\documentclass[a4paper,fleqn,usenatbib]{mnras}

\usepackage{newtxtext,newtxmath}
\usepackage[T1]{fontenc}
\usepackage{ae,aecompl}

\usepackage{graphicx}	
\usepackage{amsmath}	
\usepackage{amssymb}	
\usepackage{makeidx}

\title[BASS XIX: Ionized outflows properties]{BAT AGN Spectroscopic Survey - XIX: Type\,1 versus Type\,2 AGN dichotomy from the point of view of ionized outflows}

\author[A.F. Rojas et al.]{
A.F. Rojas$^{1,2}$
E. Sani$^{1}$,
I. Gavignaud$^{2}$,
C. Ricci$^{3,4}$,
I. Lamperti$^{5}$,
M. Koss$^{6}$
\newauthor 
B. Trakhtenbrot$^{7,8}$, 
K. Schawinski$^{8}$,
K. Oh$^{9}$, 
F.E. Bauer$^{10,11,12}$,
M. Bischetti$^{13}$,
\newauthor 
R. Boissay-Malaquin$^{14}$
A. Bongiorno$^{13}$,
F. Harrison$^{15}$, 
D. Kakkad$^{1}$,
N. Masetti$^{16,2}$,
\newauthor 
F. Ricci$^{10}$,
T. Shimizu$^{17}$, 
M. Stalevski$^{18,19}$, 
D. Stern$^{20}$,
and G. Vietri$^{21,22}$
\\
$^{1}$ European Southern Observatory, Alonso de Cordova 3107, Casilla 19, Santiago 19001, Chile\\
$^{2}$ Departamento de Ciencias Fisicas, Universidad Andres Bello, Campus La Casona, Fernandez Concha 700, Santiago, Chile\\
$^{3}$ N\'ucleo de Astronom\'ia de la Facultad de Ingenier\'ia, Universidad Diego Portales, Av. Ej\'ercito Libertador 441, Santiago, Chile\\
$^{4}$ Kavli Institute for Astronomy and Astrophysics, Peking University, Beijing 100871, People's Republic of China\\
$^{5}$ Department of Physics and Astronomy, University College London, Gower Street, London WC1E 6BT, UK \\
$^{6}$ Eureka Scientific, Inc., 2452 Delmer Street Suite 100, Oakland, CA 94602-3017, USA\\
$^{7}$ The Raymond and Beverly Sackler School of Physics and Astronomy, Tel Aviv University, Tel Aviv 69978, Israel \\
$^{8}$ Institute for Particle Physics and Astrophysics, ETH Z{\"u}rich, Wolfgang-Pauli-Strasse 27, CH-8093 Z{\"u}rich, Switzerland\\
$^{9}$ Department of Astronomy, Kyoto University, Oiwake-cho, Sakyo-ku, Kyoto 606-8502, Japan\\
$^{10}$ Instituto de Astrof\'isica, Facultad de F\'isica, Pontificia Universidad Cat\'olica de Chile, Casilla 306, Santiago 22, Chile \\
$^{11}$ Millennium Institute of Astrophysics (MAS), Nuncio Monse{\~{n}}or S{\'{o}}tero Sanz 100, Providencia, Santiago, Chile \\
$^{12}$ Space Science Institute, 4750 Walnut Street, Suite 205, Boulder, Colorado 80301 \\
$^{13}$ INAF - Osservatorio Astronomico di Roma, Via Frascati 33, 00078 Monte Porzio Catone (Roma), Italy \\
$^{14}$ Massachusetts Institute of Technology, Kavli Institute for Astrophysics, Cambridge, MA 02139, USA \\
$^{15}$ Cahill Center for Astronomy and Astrophysics, California Institute of Technology, Pasadena, CA 91125, USA \\
$^{16}$ INAF - Osservatorio di Astrofisica e Scienza dello Spazio, via Gobetti 93/3, I-40129, Bologna, Italy \\
$^{17}$ Max-Planck-Institut f{\"u}r extraterrestrische Physik, Postfach 1312, D-85741 Garching, Germany \\
$^{18}$ Astronomical Observatory, Volgina 7, 11060 Belgrade, Serbia \\
$^{19}$ Sterrenkundig Observatorium, Universiteit Gent, Krijgslaan 281-S9, Gent B-9000, Belgium \\
$^{20}$ Jet Propulsion Laboratory, California Institute of Technology, 4800 Oak Grove Drive, MS 169-224, Pasadena, CA 91109, USA  \\
$^{21}$ Cluster of Excellence, Boltzmann-Str. 2, 85748 Garching bei M{\"u}nchen, Germany \\
$^{22}$ European Southern Observatory, Karl Schwarzschild Str 2, 85748 Garching, Germany \\
}

\date{Accepted 2019 November 27. Received 2019 October 30; in original form 2019 August 5.}

\pubyear{2019}

\begin{document}
\label{firstpage}
\pagerange{\pageref{firstpage}--\pageref{lastpage}}

\maketitle

\begin{abstract}
We present a detailed study of ionized outflows in a large sample of $\sim$650 hard X-ray detected AGN. Using optical spectroscopy from the BAT AGN Spectroscopic Survey (BASS) we are able to reveal the faint wings of the [O\,{\sc iii}] emission lines associated with outflows covering, for the first time, an unexplored range of low AGN bolometric luminosity at low redshift ($z\sim$0.05).  We test if and how the  incidence and velocity of ionized outflow is related to AGN physical parameters: black hole mass ($\rm M_{BH}$), gas column density ($\rm N_{H}$), Eddington Ratio ($\rm \lambda _{Edd}$), [O\,{\sc iii}], X-ray, and bolometric luminosities. We find a higher occurrence of ionized outflows in type\,1.9 (55\%) and type\,1 AGN (46\%) with respect to type\,2 AGN (24\%). While outflows in type\,2 AGN are evenly balanced between blue and red velocity offsets with respect to the [O\,{\sc iii}] narrow component, they are almost exclusively blueshifted in type\,1 and type\,1.9 AGN. 
We observe a significant dependence between the outflow occurrence and accretion rate, which becomes relevant at high Eddington ratios (Log($\rm \lambda_{Edd}$) $\gtrsim -1.7$). We interpret such behaviour in the framework of covering factor-Eddington ratio dependence.  
We don't find strong trends of the outflow maximum velocity with AGN physical parameters, as an increase with bolometric luminosity can be only identified when including samples of AGN at high luminosity and high redshift taken from literature.

\end{abstract}




\section{Introduction}

A direct link between galaxy formation and the growth of the central supermassive black hole (SMBH) was found almost 20 years ago with the discovery of black hole (BH) mass--bulge scaling relations (e.g. \cite{Ferrarese2000, Gebhardt2000,Sani2011,Kormendy2013}). The underlying mechanism linking the two processes is thought to be Active Galactic Nuclei (AGN) feedback, i.e. the release of energy by the AGN into the interstellar medium (ISM). 
Theoretical models have proposed that AGN feedback regulates star formation (SF) by removing and/or heating the gas, so that a star-forming galaxy evolves into a red early-type galaxy (e.g. \cite{Granato2004,DiMatteo2005,Hopkins2006,Merloni2008}). The majority of AGN feedback is thought to come in two flavours, at low Eddington ratios radio-jets dominate kinetic feedback (e.g. \cite{Fabian2012}) whereas at higher Eddington ratios radiation pressure from the accretion disk is responsible for gas outflows. In a considerable fraction of AGN there are signatures of outflows capable of removing significant amounts of cold gas from the host galaxy \citep{Rupke2013,Feruglio2010,Feruglio2015,Cicone2014} or heating the gas (e.g. \cite{Shangguan2018}). 
Nonetheless, studies exploring the role of AGN with respect to SF have led to ambiguous results, not finding any compelling evidence of negative feedback \citep{Balmaverde2016} and even evidence of positive feedback, with outflows generating local SF \citep{Ishibashi2012, Cresci2015, Maiolino2017}. Additionally, some authors showed that the observed SMBH-host scaling relations can be reproduced without invoking feedback mechanisms, simply as a result of the hierarchical assembly of BHs and stellar mass through galaxy merging (e.g. \cite{Peng2007, Hirschmann2010, Jahnke2011}).


High velocity ($>$ 1000 km\,s$^{-1}$) and extended AGN-driven outflows are frequently detected in local and high-redshift galaxies, at different luminosities, in ionized, neutral and molecular gas. 
In particular, high resolution, integral field spectroscopic observations and millimetre interferometers have revealed the presence of ionized and molecular gas outflows, respectively, with high velocities, $>$ 1000 $\rm km\,s^{-1}$, in low and high redshift galaxies, and can allow the characterization of their outflow properties, such as the extension of the ejected material as well as the entrained masses and associated energy \citep{Harrison2012, Harrison2014, Harrison2016, Cicone2014, Cresci2015, Perna2015a,Perna2015b, Brusa2016, Carniani2016, Kakkad2016, Bischetti2017, Kang2018}.

Numerous efforts have been done to test the occurrence of ionized outflows, finding them in a large fraction of SDSS AGN, from $\sim$20-40\% to $\sim$50-70\% depending on the AGN type \citep{Balmaverde2016,Harrison2016, Perna2017a, Veroncetty2001, Woo2016, Woo2017, Rakshit2018}. 

However, most of the samples considered by these studies are incomplete due to biases against absorption in the optical/soft X-ray band. Therefore it has been difficult to place the outflow signatures of galaxy populations in the context of both obscured and unobscured AGN. 
The physical processes responsible for the origin of outflows, their frequency within an unbiased sample of AGN, and how they can affect the evolution of the host galaxy and its ISM, are still open questions. Hard X-ray selection offers the least biased AGN selection (e.g. Baumgartner et al. 2013) thanks to the low contamination from other sources within the host galaxy and the high penetration ability of hard X-rays up to large gas column densities ($N_{\rm H} \le \sigma_{\rm T}^{-1} \eqsim  1.5 \times 10^{24} \rm cm^{-2}$, where $\sigma_{\rm T}$ is the Thomson cross section), see e.g. \citet{Ricci2015}. The \textit{BAT AGN Spectroscopic Survey} (BASS; \cite{Koss2017, Ricci2017d}) allows us to take advantage of an unbiased local sample of $\sim$ 650 AGN to study the occurrence of ionized outflows traced by the [O\,{\sc iii}]$\lambda$4959,5007 emission lines and how they are related to other key AGN properties (bolometric luminosity $L_{\rm Bol}$, Eddington ratio $\lambda_{\rm Edd}$\,= $L_{\rm Bol}$/$L_{\rm Edd}$ \,$\propto$ $L_{\rm Bol}$/$M_{\rm BH}$, column density $N_{\rm H}$, intrinsic $L_{\rm X}$). The [O\,{\sc iii}] emission lines are frequently used to study outflows because they are a bright doublet in the optical range. They trace the gas ionized by the AGN at distances which are not affected by the gravitational potential of the central SMBH. Any broadening of these lines can be safely assumed to be due to gas kinematics in the narrow-line region (NLR). 

Moreover, studies exploring the role of radiation pressure on the obscuring material lead to interesting results which can be tested with our study on the incidence of outflows. \citet{Ricci2017c} investigated the obscuration properties of the BASS sample and found that radiative feedback on dusty gas regulates the distribution of the obscuring material, 
such that SMBHs accreting with Log($\lambda _{Edd}$) $>$ -1.5 have a lower covering factor ($\sim$ 40\%). This is likely due to the fact that radiation pressure is able to expel a large fraction of the dusty gas in the form of outflows.

The aim of our work is to test such a hypothesis by looking at the incidence of ionized outflows and how they relate, in terms of wind velocities, outflowing mass rate, and wind power, to the central SMBH and host galaxy parameters. The paper is structured as follows: in Section 2, we provide information about the sample selection; in Section 3, we describe the methods adopted to determine the outflows velocities from the optical spectra and fitting method used; in Section 4, we present the results derived from our multi-component line fits to the [O\,{\sc iii}]$\lambda$5007 profiles; in Section 5, we discuss how the occurrence of outflows evolves as a function of parameters such as the Eddington ratio, [O\,{\sc iii}]$\lambda$5007 luminosity and spectral classification (i.e. optical type\,1, type\,2); finally, in Section 6 we summarize the main conclusions of our work. In the following, we assume a cosmology with $H_{0}$ = 70 km $s^{-1}Mpc^{-1}$, $\Omega _{\Lambda}$ = 0.7, and $\Omega _{m}$ = 0.3. 

\section{Data}

The Burst Alert Telescope (BAT) instrument onboard \textit{Swift} satellite \citep{Gehrels2004} is carrying out an all-sky survey in the ultra-hard X-ray band ($>$ 10 keV) that, as of the first 70 months of operation, identified 1210 objects \citep{Baumgartner2013} of which 836 are classified as AGN based on their associations with objects in the medium and soft energy X-ray band, and using NED and SIMBAD databases. 

The optical spectral data used in this work are part of the BASS sample \citep{Koss2017}, a dedicated multi-wavelength follow-up project for the nearby, powerful AGN identified by \textit{Swift}-BAT 70 month catalog. BASS optical data release 1 (DR1, \cite{Koss2017})\footnote{https://www.bass-survey.com} compiled optical spectra of 642 AGN, taken from public surveys such as SDSS or previously published papers (67\%, 433/642, e.g. \cite{Rojas2017, Masetti2013}, and references therein), and from targeted campaigns (33\%, 209/642). 
DR1 presents redshift, classification, black hole mass ($M_{\rm BH}$), bolometric luminosity ($L_{\rm Bol}$), Eddington ratio ($\lambda_{\rm Edd}$) and optical spectral properties such as emission line strengths and velocity dispersions for the majority of obscured and unobscured AGN (74\%, 473/642), including 340 AGN for which $M_{\rm BH}$ and $\lambda_{\rm Edd}$ are measured for the first time. 
 
The BASS sample is nearly unbiased against obscuration up to Compton-thick levels ($N_{\rm H}$ > 10$^{24}$ cm$^{-2}$) due to its selection from the hard X$-$ray band (14--195 keV). \citet{Koss2017} presented an overview of the optical spectroscopic data. Additional extended multi-wavelength campaigns from near-IR (NIR) to soft X-ray wavelengths \citep{Berney2015, Lamperti2017, Oh2017, Ricci2017c, Trakhtenbrot2017, Ricci2017d, Oh2018, Powell2018, Ricci2018, Shimizu2018} have allowed us to further characterize the BASS sample, studying for example the connection between X-ray and optical obscuration, NIR lines, X-ray photon index, absorption and coronal properties.



\subsection{Sample Selection}


Our main focus is to identify outflows, and therefore to spot the faint wings of the [O\,{\sc iii}]$\lambda$5007\,\AA\ emission line and test how asymmetric/broad profiles associated with ionized outflows relate to other key AGN properties. To provide reliable measurements, we excluded from the analysis cases where the [O\,{\sc iii}] emission is too faint to properly parametrize the wings or where the noise on the continuum is so high that it dilutes such wings. Therefore, we adopt a threshold in S/N to discard unreliable spectra, where the noise level was estimated by considering the dispersion of the continuum in the [O\,{\sc iii}] region, and the signal from the [O\,{\sc iii}] emission. From the 642 DR1 spectra, we exclude 59 sources with S/N $<$ 7, and 36 spectra that do not cover the H$\beta$ + [O\,{\sc iii}] region. In total, we consider 547 spectra of the whole sample (85\% of the BASS DR1 catalogue). After these cuts, our sample contains 210 type\,2 (defined as AGN with only narrow permitted and forbidden emission lines in the optical spectra, full width at half-maximum or FWHM $\leq$ 1000 km\,s$^{-1}$), 92 type\,1.9 (defined as AGN with only narrow emission lines aside from a broad profile of FWHM $\ge$ 1000 km\,s$^{-1}$ seen from H$\alpha$) and 245 "type\,1" (defined as AGN with broad permitted lines beyond H$\alpha$ plus narrower forbidden lines; note that we group together all Seyfert subtypes 1.0, 1.2, 1.5, and 1.8 here).
For the following analysis, we adopt the AGN classification determined in \citet{Koss2017} by means of BPT (Baldwin, Phillips \& Terlevich) diagnostic diagrams, depending on the presence and strength of broad emission lines, and the type\,1 subtypes classification based on \citet{Winkler1992}.

\section{Spectral Analysis} 

In order to identify outflow signatures in our sample, 
we use the profile of [O\,{\sc iii}]$\lambda$5007\,\AA\ emission lines in the rest-frame optical spectra. 

These emission lines are produced through a forbidden transition and are only emitted by low-density gas ($n_{\rm H}\ \leq\ 10^{6}\ cm^{-3}$) located in the NLR, which is an extended region (1--1000 pc) of gas clouds photoionized by the non-stellar continuum emission of the AGN \citep{book:peterson}. The velocity dispersions of such clouds are intrinsically narrow (FWHM $\leq$ 1000 km\,s$^{-1}$) and therefore any broadening or shifting of the [O\,{\sc iii}] emission lines is interpreted as the result of outflowing gas in the NLR, which can be extended over kpc scales (e.g., \citealt{Pogge1989}), as opposed to the more dense, sub-pc-sized broad-line region (BLR).




The emission-line profiles of ionized gas in AGN host galaxies often appear to be non-Gaussian and asymmetric, composed of a narrow (systemic) component plus a wing. In addition, the line centroids are often blueshifted (or redshifted) relative to the systemic velocity of the host galaxy, as measured by the stellar light. Such asymmetries are usually interpreted as a result of outflowing gas in the NLR. The effect tends to be stronger in the higher ionization lines \citep{Veilleux1991}, and thus we focus our attention on the [O\,{\sc iii}]$\lambda$4959,5007\,\AA\ doublet, and in particular on the [O\,{\sc iii}]$\lambda$5007\,\AA\ emission line. 

In this section we describe the procedure used to model the profile of the H$\beta$ + [O\,{\sc iii}] spectral region (4400--5500\,\AA).

\begin{figure*}
\centering
\includegraphics[width=0.45\textwidth]{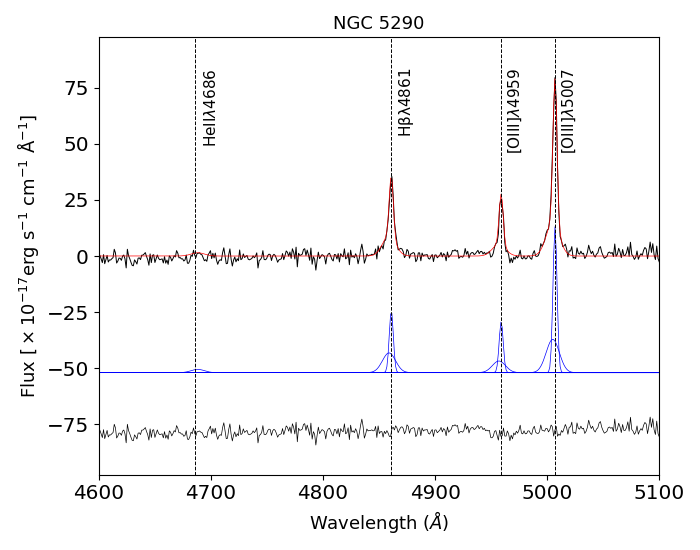}
\includegraphics[width=0.45\textwidth]{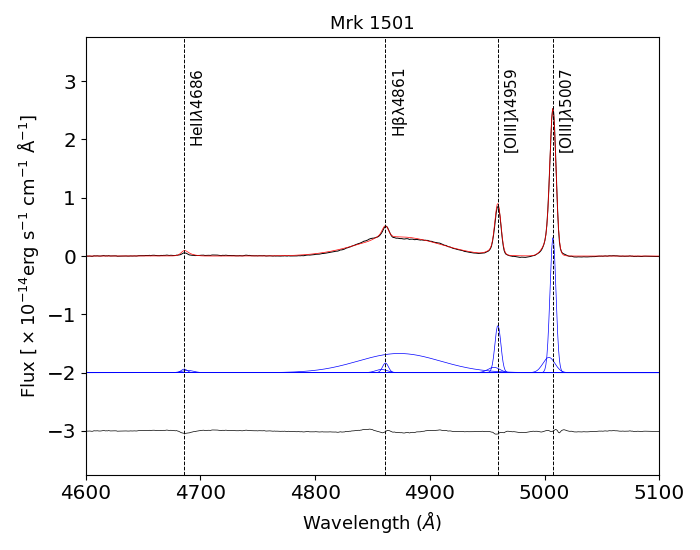}
\caption{Examples of our fitting method. {\it Left}: An example of type\,2 AGN fitting after stellar component subtraction. {\it Right}: Spectral fitting of a type\,1 AGN after continuum and Fe\,{\sc ii} subtraction. For each object, we show the spectra (in black), the components fitted for emission lines (in blue), the overall best-fit model (in red), and the residuals below. The dashed vertical lines mark the location of He\,{\sc ii}$\lambda$4686, H$_{\beta} \lambda$4861 and [O\,{\sc iii}]$\lambda$4959, 5007.}
\label{fig:spec_fitting}
\end{figure*}

\subsection{Continuum fitting (4400--5500\,\AA)}


To model the inner regions of the AGN and obtain the parameters characterizing the gas kinematics, we adopt a multi-component spectral fit using {\sc PySpeckit}, an extensive spectroscopic analysis toolkit for astronomy, which uses a Levenberg-Marquardt minimization algorithm \citep{Ginsburg2011}. 


Prior to the multi-component fitting of the emission lines, we first removed the underlying AGN continuum or the host galaxy emission. A pseudo-continuum slope is define by two narrow (10\,\AA) emission-line free regions (4440--4450\,\AA\  or 4720--4730\,\AA, and 5110--5120\,\AA; see \citet{Shen2016}). Depending on the AGN spectral type and complexity, we additionally used a Fe\,{\sc ii} grid of templates or stellar population synthesis models to fit the underlying continuum (see below for details). 
The best fitting template is chosen by minimizing the $\chi^2$, and it is then subtracted from the spectra so that the pseudo-continuum made by the Fe\,{\sc ii} emission can be measured to fit the emission lines. 

A major complication in fitting type\,1 AGN spectra is that the complex Fe\,{\sc ii} emission lines contaminate almost entirely the 4400-5500\,\AA\ waveband including the [O\,{\sc iii}] profile. 
We used both empirical and synthetic Fe\,{\sc ii} templates to fit the iron emission \citep{Boroson1992, Bischetti2017, Vietri2018}. However, we were not able to obtain a satisfactory fit to the Fe\,{\sc ii} emission using the empirical ones because the [O\,{\sc iii}] emission lines in their spectra are blended with the iron features thus producing artificial outflow signals. Therefore, we considered 20 synthetic models to account for the Fe\,{\sc ii} emission in the H$\beta$ + [O\,{\sc iii}] region and clean our type\,1 AGN sample of that emission. The templates were created by the photoionization simulation code {\sc cloudy} \citep{Ferland2013}, for different ionizing photon fluxes emitted by the primary source [$\Phi$(H)] and electron densities in the BLR clouds ($n_{\rm e}$), with and without the possibility of a microturbulence velocity ($u_{\rm turb} {=} 0$ or 100 km\,s$^{-1}$). The parameters for each Fe\,{\sc ii} model are listed in Table~\ref{tab:table_FeII}.
Each template is then convolved with a set of Gaussian profiles of increasing width from 1200\,km\,s$^{-1}$ to 20000\,km\,s$^{-1}$ in steps of 25\,km\,s$^{-1}$. We refer hereafter to the resulting broadened templates as "sub-templates".

The continuum fitting procedure applied to each type\,1 AGN spectrum is as follows:\\
- A linear pseudo-continuum is fitted within the spectral windows 5110--5120\,\AA\ and 4400--4450 \,\AA\ (or 4720--4730\,\AA) and subtracted from the spectrum.\\
- For each AGN spectrum, a best-fit Fe\,{\sc II} sub-template is selected from each of the 20 original synthetic models, restricting the broadening width to be within 2000\,\AA\ of the FWHM of H$\beta$, that is generally what we observed (Table 9 of the BASS DR1; \citealt{Koss2017}). The Fe\,{\sc II} sub-templates are normalized considering the AGN spectrum within the 4450--4750\,\AA\ and 5050--5400\,\AA\ windows. The selected sub-template is the one which minimizes the residuals in the 4450-4650 \,\AA\ and 5100-5400 \,\AA\ spectral regions where the iron emission is usually strongest. \\
- Finally, we select the normalized best-fitting Fe\,{\sc II} sub-template with the lowest residuals from the type\,1 AGN spectra.  Our final sample with a satisfactory continuum and Fe\,{\sc II} template subtraction totals 167 type\,1 AGN.

In the case of type\,2 and 1.9 AGN, the major contamination in the spectra is the stellar component. A large set of single stellar population synthesis templates was used to model and subtract the host stellar component from each galaxy spectrum using the penalized PiXel Fitting software (pPXF; \citealt{Cappellari2004}). The templates used are from the Miles Indo-U.S. Catalog (MIUSCAT) library of stellar spectra \citep{Vazdekis2012}. The galaxy continuum and stellar absorption features were removed, as explained in \citet{Koss2017} and in \citet{Lamperti2017}. 
Our final sample is comprised of 210 type\,2 AGN, and 92 type\,1.9 AGN with a satisfactory host continuum subtraction. 


Finally, depending on the spectral type and gas kinematics, the emission lines are modeled by means of multi-Gaussian components (narrow, broad and outflow) in the continuum and Fe\,{\sc ii} subtracted spectra for type\,1 AGN or the host continuum subtracted spectra of type\,2 and type\,1.9 AGN, respectively. 

\begin{table}
\caption[]{Parameters of the Fe\,{\sc ii} simulated templates for different physical conditions of the BLR. Each column indicates a template, with the ionized photon flux emitted by the primary source [$\Phi (H)$] and electron density in the BLR clouds ($n_{\rm e}$) used as input for {\sc cloudy}. For these conditions, we consider the case with a microturbulence velocity and the case without that velocity.}
\label{tab:table_FeII}
\setlength{\tabcolsep}{5pt}
\begin{center}
\begin{tabular}{ccccccccccc}
\noalign{\smallskip}
\hline
\hline
\noalign{\smallskip}
\multicolumn{1}{c}{log($\frac{\Phi (H)}{{\rm cm}^{-2}{\rm s}^{-1}})$: }  & 17 & 17 & 19 & 17 & 19 & 21 & 17 & 19 & 21 & 23  \\
\hline
\noalign{\smallskip}
\multicolumn{1}{c}{log($\frac{n_{\rm e}}{{\rm cm}^{-3}})$:} & 8 & 10 & 10 & 12 & 12 & 12 & 14 & 14 & 14 & 14  \\
\hline
\end{tabular}
\end{center}
\end{table}

\subsection{Multi-component line fitting (4400--5500\,\AA)}


\subsubsection{\textbf{Seyfert 1.9 and Seyfert 2 AGN}}

We fit a linear (pseudo-) continuum based on continuum windows (4660--4670\,\AA, 4700--4750\,\AA, and 5040--5200\,\AA). Seyfert 1.9 AGN are characterized by a broad component detected in H$\alpha$ but not H$\beta$; therefore we describe the emission line fitting of Seyfert 1.9 and 2 AGN together. We fit narrow components for the He\,{\sc ii}$\lambda$4686, H$\beta$ and [O\,{\sc iii}]$\lambda$4959,5007 emission lines, and we consider an additional broad outflowing (offset) component for H$\beta$ and for each [O\,{\sc iii}] line in order to distinguish outflow signatures. We note that a single-Gaussian fit (i.e. one per emission line) often fails to account for the complex emission-line profiles seen in many BASS AGN, and a second broad component fit to the [O\,{\sc iii}] lines allows us to find and characterize asymmetric profiles.

 
We tied together the rest-frame centroids of the narrow components of the emission lines ($\lambda_{\rm He\,II} {=} \lambda_{\rm H\beta} {-} 175$\,\AA ; $\lambda_{\rm H\beta} {=} \lambda_{\rm [O\,III]5007} {-} 146$\,\AA ; $\lambda_{\rm [O\,III]4959} {=} \lambda_{\rm [O\,III]5007} {-} 48$\,\AA) and the broad components of the lines were tied together in order to identify their blue- or redshift common to all these lines. 

We allow for velocity offsets of up to $\pm$ 600 km\,s$^{-1}$ in the narrow component and $\pm$ 1800 km\,s$^{-1}$ in the broad components, compared to the database redshift. Such large velocity shifts are motivated by the observations of a mean velocity shift of the broad H$\beta$ to the systemic redshift of 109 km\,s$^{-1}$ with a scatter of 400 km\,s$^{-1}$ in a sample of 849 quasars \citep{Shen2016}.  


The widths of the narrow lines are also tied together ($\sigma_{\rm H\beta}$ (narrow) =  $\sigma_{\rm [O\,III]5007}$ (narrow) = $\sigma_{\rm [O\,III]4959}$ (narrow)), with an initial input guess of $\sigma_{\rm narrow}$ = 300 km\,s$^{-1}$ ($\sim$5\,\AA). We constrain the FWHM of the narrow lines to be less than 1200 km\,s$^{-1}$ in all cases. 
Likewise, we tied together the widths of the broad components ($\sigma_{\rm H\beta}$ (broad) = $\sigma_{\rm [O\,III]5007}$ (broad) = $\sigma_{\rm [O\,III]4959}$ (broad)) and set an initial value of $\sigma_{\rm broad}$ = 500 km\,s$^{-1}$ ($\sim$8\,\AA) with no upper limit.

For an initial estimate of the amplitudes of the narrow lines, we measured the maximum value of the (continuum-subtracted) flux density in the region where the line is supposed to be ($\pm$5\,\AA\ of theoretical wavelength), and used that value as the initial guess. For the amplitude of the broad components, we used 50\% of the maximum value as the initial guess. 
Additionally, we fixed the intensities of [O\,{\sc iii}]$\lambda$4959 and [O\,{\sc iii}]$\lambda$5007 to the theoretical ratio 1:2.86 for both broad and narrow components of the [O\,{\sc iii}] doublet, according to atomic physics \citep{Storey2000, Dimitrijevic2007}.


\subsubsection{\textbf{Seyfert 1 AGN}} 

After the continuum and Fe\,{\sc ii} emission have been subtracted, we consider two velocity components (narrow+broad) for each line in the region: He\,{\sc ii}$\lambda$4686, H$\beta \lambda$4861, [O\,{\sc iii}]$\lambda$4959,5007, adopting the same initial guesses for amplitudes, wavelengths and widths used for type\,2 and type\,1.9 AGN. But in these AGN, we consider an additional broad component of H$\beta$ to account for the BLR emission. Therefore, we have three components for H$\beta$, one narrow component, one broad component for the outflow plus a very broad BLR component. The first broad components is modeled using the same initial guesses as for type\,2 AGN in order to model a possible outflow component, while the second broad component is left free in wavelength and width because we do not know the characteristics of the BLR for each AGN.  Fig.~\ref{fig:spec_fitting}, we show two examples of spectra illustrating our fitting method. 

\subsubsection{\textbf{Errors estimation}}

For all the sources, we estimated the errors associated with each fitted parameter using Monte Carlo simulations. We repeated the entire fitting procedure (including line emission components and continuum) 100 times, each time adding an amount of noise to each spectral bin randomly drawn from a normal distribution based on the standard deviation of the corresponding local continuum level of each spectrum. From these, we computed the standard deviation of the mean of the 100 measurements and used this value as an estimate of the error at the 1-$\sigma$ confidence level. We performed a visual inspection of all the emission line fits to verify proper fitting. We stress that with this method we can address any kind of uncertainties, included those related to the continuum fitting to which low amplitude broad line components can be highly sensitive.


\section{Results}

Here we describe the strategy to detect outflows in the BASS DR1 sample and their incidence.  Outflow velocities and related properties were estimated from the [O\,{\sc iii}]$\lambda$5007\,\AA\ emission lines adopting well-known parametric and non-parametric prescriptions, as presented below. These two prescriptions serve as consistency checks and allow us to properly compare our results with various literature samples (e.g., \citealt{Alexander2012, Harrison2014}).

\subsection{Detection of ionized outflows}

We used the fitted parameters of the two components of [O\,{\sc iii}]$\lambda$5007\,\AA\ emission line (narrow and broad) to define an outflow detection and corresponding velocity. 

A non-Gaussian profile in the [O\,{\sc iii}] emission lines, with relatively stronger wings/tails than a Gaussian, is indicative of an ionized wind. The profiles usually show a blue (or occasionally red) wing, i.e. the profile of the lines is reproduced with two components, a narrow one associated with the emission from the NLR and the second broad component that can be shifted. In this work this second component is considered to be an outflow candidate. 


To consider a secondary broad component as reliable, we imposed a detection limit of 3$\sigma$, based on the standard deviation of the baseline fitting without emission lines. 
Since most line profiles are found to be asymmetric \citep{Perna2017a}, we consider a blue or redshifted outflow detection when the wavelength shift between the two components of [O\,{\sc iii}] line is significant compared to the errors.

\begin{figure}
\includegraphics[width=\columnwidth]{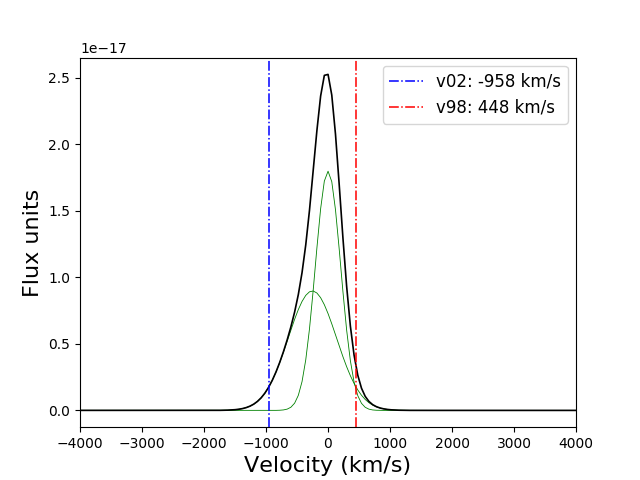}
\vspace{-0.5cm}
\caption{Example of [O\,{\sc iii}] emission-line reconstruction from the narrow and broad components obtained after fitting 100 Monte Carlo simulations. 
The non-parametric 98th (v98) and 2nd (v02) velocity percentiles are shown as vertical dashed lines, while their values are quoted with respect to the narrow centroid as 0\,km\,s$^{-1}$ in the legend; the larger of the absolute values of v02 and v98 is considered v$_{\rm max}$ (in this case v02).
}
\label{fig:nonParamMethod}
\end{figure}


\begin{figure}
\includegraphics[width=\columnwidth]{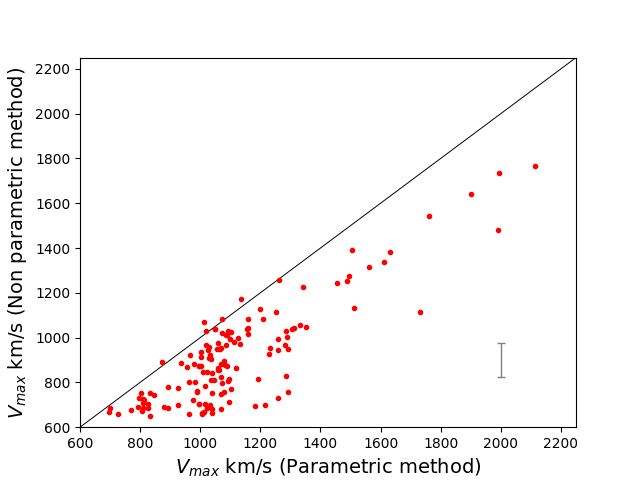}
\vspace{-0.5cm}
\caption{Comparison between maximum outflow velocities estimated using the two detection methods, parametric (x-axis) and non-parametric (y-axis). A typical error bar ($\rm \pm 150 km/s$) is shown for reference.}
\label{fig:veloc_comparison}
\end{figure}

We define the wavelength shift between the broad component that represents the outflow and the narrow component of the line as: 

\begin{equation}
\Delta \lambda = \lambda _{\rm [O\,III], Narrow} - \lambda _{\rm [O\,III], Broad}.
\end{equation} \\

\noindent A blueshifted outflow arises when $\Delta \lambda$ > $\epsilon _{\lambda}$ and a redshifted outflow when $\Delta \lambda$ < -$\epsilon _{\lambda}$, where $\epsilon _{\lambda}$ = $\sqrt{(\epsilon _{\rm broad} ^{2} + \epsilon _{\rm narrow} ^{2})}$ with $\epsilon _{\rm broad}$ and $\epsilon _{\rm narrow}$ being the fitting errors on the central wavelengths of both components.

Finally, the maximum outflow velocity was estimated following the approach of \citet{Rupke2013}, assuming that the outflow expands with constant velocity: 

\begin{equation}
{\rm v}_{\rm max} = \Delta \lambda + 2\sigma_{\rm [O\,III], Broad} 
\end{equation} \\

\noindent where $\sigma_{\rm [O\,III], Broad}$ is the velocity dispersion parameter of the Gaussian representing the broad component of the [O\,{\sc iii}] line. 

In order to avoid biases associated with the outflow detection criteria assumed, we also adopt a non-parametric velocity estimator frequently used in literature (see, e.g., \cite{Zakamska2014, Harrison2014, Balmaverde2016, Perna2017a}) to estimate the outflow properties from the [O\,{\sc iii}] emission line. To start, we reconstruct a synthetic line profile for [O\,{\sc iii}]$\lambda$5007\,\AA\ using the average of the best-fitted parameters from in the 100 Monte Carlo simulations discussed in Sect. 3.2. 
Then, we measure the velocity at which a given fraction of the line flux is collected using the cumulative flux function $F({\rm v}) {=} \int_{-\infty}^{\rm v} F_{\rm v}({\rm v}')d{\rm v}'$ and defining different percentiles of the overall line flux (v02, v05, v50, etc). An example can be found in Fig.~\ref{fig:nonParamMethod}.

The degree of line asymmetry is estimated by the dimensionless $R$ parameter introduced by \citet{Zakamska2014}: 

\begin{equation}
R = \frac{\rm (v95-v50)-(v50-v05)}{\rm (v95-v05)}.
\end{equation} \\

\noindent When $R {<} 0$, a blue prominent broad wing is present and the maximum outflow velocity ${\rm v}_{\rm max}$ is defined as v02. When $R {>} 0$, a red wing is present and the ${\rm v}_{\rm max}$ corresponds to v98.
The velocity offset of the broad wing is defined as $\Delta {\rm v} {=} \frac{\rm (v05+v95)}{2}$ and the velocity shift respect to the velocity peak (v$_{\rm p}$=v50) of the whole line is given by $|\Delta {\rm v}| {-} {\rm v}_{\rm p}$.

\begin{figure}
\includegraphics[width=\columnwidth]{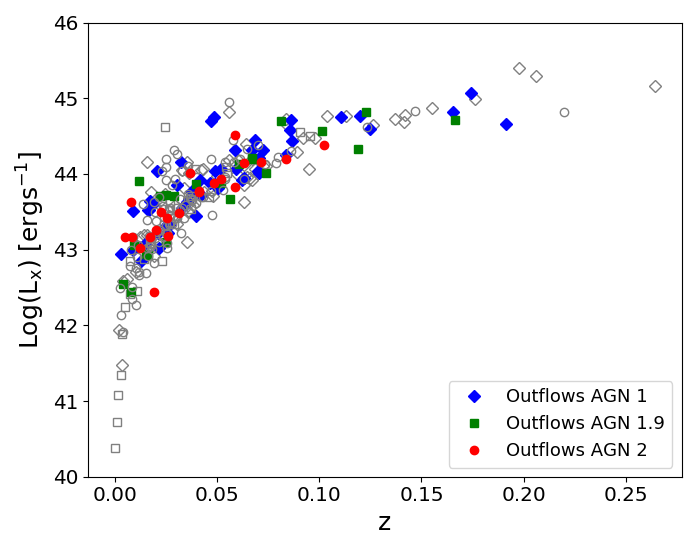}
\vspace{-0.5cm}
\caption{X-ray luminosity (14--195\,keV) as a function of redshift for our sample, with symbols distinguishing: AGN\,1 (diamonds), AGN\,1.9 (squares) and AGN\,2 (circles). Open grey symbols indicate AGN with no measurable outflow signal for the different types, while filled, color-coded symbols denote AGN with outflows detected as described in Sect. 4.1.}
\label{fig:Lx_z}
\end{figure}

\begin{figure}
\includegraphics[width=\columnwidth]{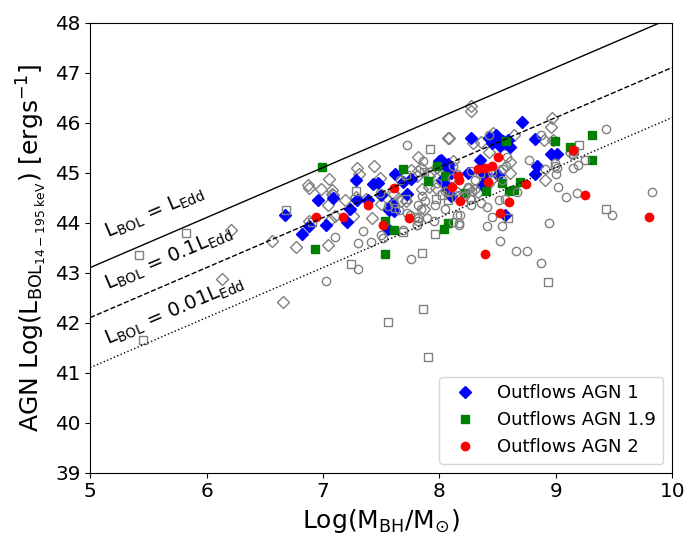}
\vspace{-0.5cm}
\caption{Bolometric luminosity as a function of SMBH mass for our sample. 
Symbols are identical to Fig.\ref{fig:Lx_z}.
The lines correspond to $L_{\rm Bol} {=} L_{\rm Edd}$ (solid), $L_{\rm Bol} {=} 0.1L_{\rm Edd}$ (dashed) and $L_{\rm Bol} {=} 0.01L_{\rm Edd}$ (dotted).}
\label{fig:Lbol_Mbh}
\end{figure}

\begin{figure*}
\centering
\includegraphics[width=0.3\textwidth]{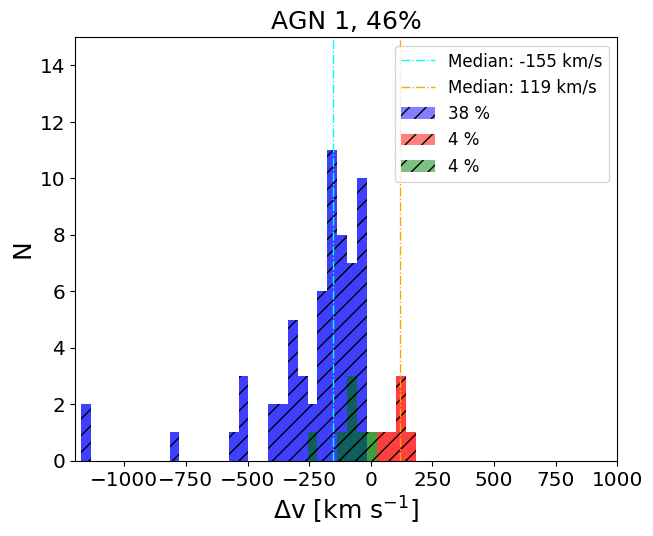}
\includegraphics[width=0.3\textwidth]{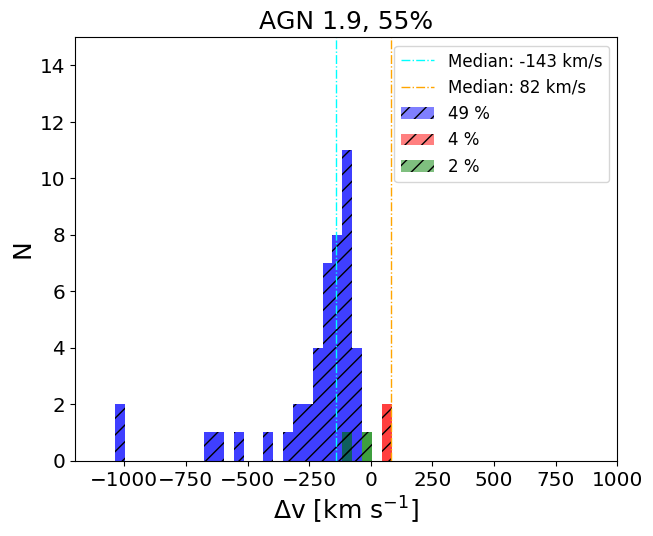}
\includegraphics[width=0.3\textwidth]{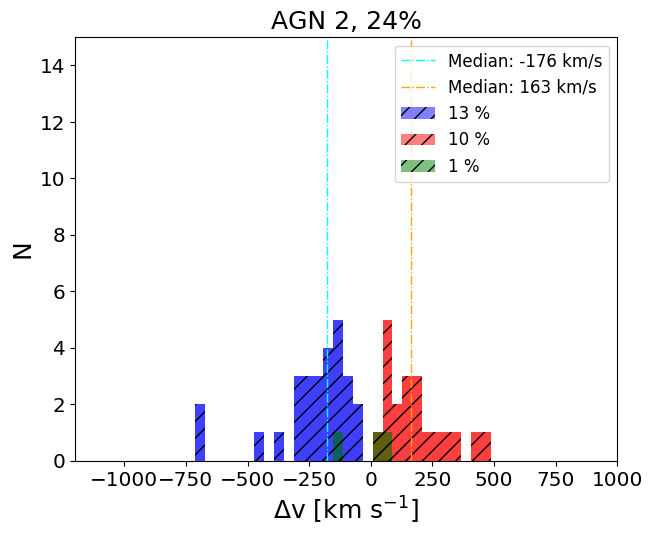} \\
\includegraphics[width=0.3\textwidth]{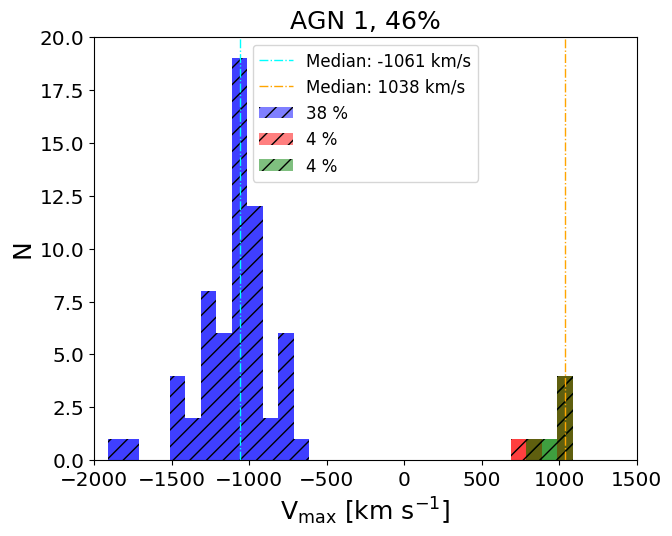}
\includegraphics[width=0.3\textwidth]{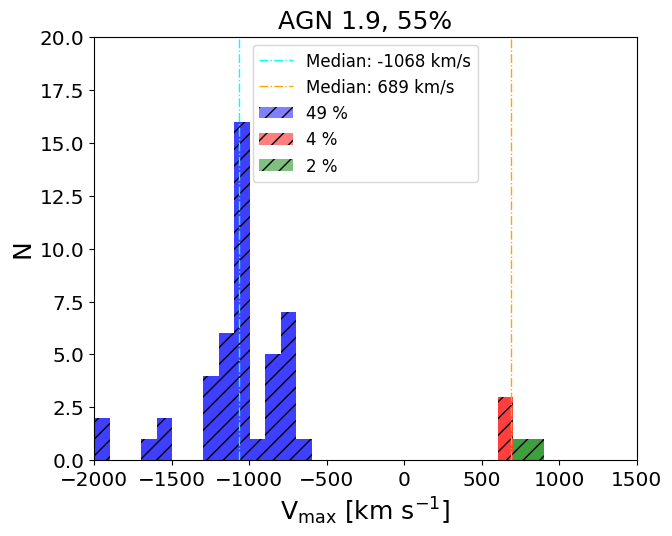}
\includegraphics[width=0.3\textwidth]{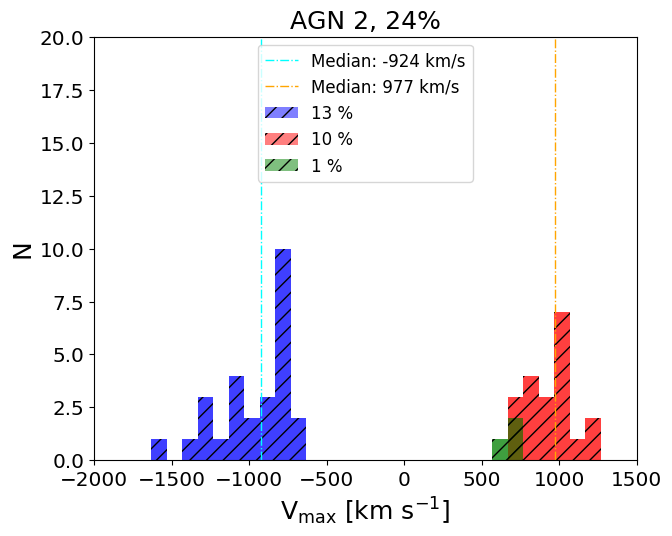}
\caption{Distributions of sources with ionized outflows as a function of wavelength shift (top) and maximum velocity (bottom). The panels show wavelength shifts and velocity distribution for approaching, receding and symmetric outflows, represented with blue, red and green histograms, respectively, for the subsamples defined on the basis of the AGN type. For each panel, the fraction of AGN with outflow is indicated, together with the fraction of receding, approaching and symmetric outflows.}
\label{fig:veloc3}
\end{figure*}


Once we have the velocities, following \citet{Perna2017a}, we consider a velocity threshold of ${\rm v}_{\rm max} {=} 650$\,km\,s$^{-1}$ to discriminate between kinematics dominated by gravitational broadening and outflow processes. This criterion is applied to the maximum velocities estimated from both the parametric and non-parametric methods. Once applied, we find consistency between the methods for the majority (65\%) of the AGN with outflow signals detected.

Overall, we considered an outflow signal to be firmly detected if the asymmetry is found by both methods. In case the outflow was found with only one method, we visually inspected the line profile to decide whether the outflows was detected reliably or not.


For the visual check, we also considered the possibility to have symmetric outflows, i.e., where the second broad component of [O\,{\sc iii}]$\lambda$5007 is necessary to reproduce the line profile, but has no considerable wavelength shift with respect to the narrow component.
With this approach we found 19 additional blueshifted, 6 redshifted outflows and 6 with signals of a symmetric outflow for type\,1 AGN. For type\,1.9, we found 9 additional blueshifted outflows, 3 additional redshifted outflows and 2 AGN with symmetric outflow. 
Finally, for type\,2 AGN, we found 11 additional blueshifted outflows, 6 redshifted and 3 with symmetric outflow after visual inspection. 


In the following, we adopt the maximum velocities of outflows estimated from the parametric method (Eq. 2). We show in Fig.~\ref{fig:veloc_comparison} a comparison between the maximum outflow velocities estimated by each method. We can see that most of the data points lie systematically below the 1:1 ratio line, with higher velocities showing a more consistent deviation. This behavior, whereby the parametric method estimates larger velocities than the non-parametric method, is expected since the parametric method uses the parameters representing the outflowing gas, while the non-parametric method estimates the velocities from the whole [O\,{\sc iii}] emission line (v02, v098), and thus its maximum velocity can be strong down-weighted by the narrow line emission.


\subsection{Outflow Statistics}

Based on the analysis described above, we detect outflows in all AGN types, spanning a broad range in luminosity and SMBH mass up to $z {\sim} 0.2$ (see Fig.~\ref{fig:Lx_z} and Fig.~\ref{fig:Lbol_Mbh}). 
Overall, 38$\pm$2\% of the BASS sample analysed (178/469) exhibit detectable ionized outflows signatures, with 29\% showing blueshifted and 7\% redshifted. 

We found that: 46$\pm$4\% of type\,1 AGN have signals of outflows, mostly blueshifted; 55$\pm$5\% of type\,1.9 AGN have outflows detections, also mostly blueshifted; and 24$\pm$3\% of type\,2 AGN have signals of outflows, but with comparable fractions being blueshifted and redshifted. 


In Fig.~\ref{fig:Lx_z}, we show AGN with outflow signals with color-coded symbols compared with AGN without outflows (open grey symbols). We find that AGN with detected outflows span X-ray luminosities between 10$^{42}$--10$^{46}$\,erg\,s$^{-1}$ for the different subtypes. Independent of AGN type, we do not find outflows for AGN with $L_{X} {<} 3\times10^{42}$\,erg\,s$^{-1}$, implying that these AGN may not be powerful enough to drive strong outflows and/or the  wings of the [O\,{\sc iii}] lines are too faint to be detected.
We also note that among the few high-luminosity AGN found with outflows at $z{\gtrsim}0.1$, the type\,2 AGN fraction is quite low (1/4). Clearly the results suffer from poor statistics, but may also arise from a dilution effect caused by the host galaxy; for a given slit width, the farther the system is, the larger is the contribution of the host galaxy entering in the slit, thus diluting a possible wind signature. Given the small number of distant sources, this potential bias anyhow does not affect our results. 

Fig.~\ref{fig:Lbol_Mbh} shows the bolometric luminosity as a function of $M_{\rm BH}$ for the AGN with outflow signals (color-coded symbols) compared with AGN without outflows (open grey symbols) in our sample. The BASS sample criteria selects a wide range of bolometric luminosities ($\sim$10$^{42}$--10$^{46}$\,erg\,s$^{-1}$) and BH masses ($\sim$10$^{7}$--10$^{10}$\,$M_{\odot}$). The lines correspond to different accretion efficiencies commonly used in the literature to estimate the fraction of AGN bolometric luminosity in the form of the outflow power. Eddington luminosities $L_{\rm Edd}$ correspond to: 1.26$\times$10$^{38}$($M_{\rm BH}$/$M_{\odot}$) erg\,s$^{-1}$. The distribution of AGN with outflows in the $L_{\rm Bol}$--$M_{\rm BH}$ plane spans a comparable range as the full BASS sample. We can see that the detected outflows generally reside at slightly higher $L_{\rm Bol}$/$L_{\rm Edd}$ values than non-detections. This might be either due to low efficiency of the AGN engine in driving the wind, or to the faintness of the wings in the [O\,{\sc iii}] line. We note however that the cut applied in S/N should mitigate this last effect. Specifically, type 1 AGN with outflows signatures are preferentially located at high Eddington ratios than type 1.9 and 2 (see Section 5.4 for more details).

The statistics of the detected outflows are shown in Fig.~\ref{fig:veloc3} for the different AGN types, where top panels indicate the wavelength shift between the narrow and broad component of the [O\,{\sc iii}] line ($\Delta \lambda$), i.e. the measured asymmetry of the line that represents the outflow itself, and the bottom panels indicate maximum outflow velocities from the geometrical models assumed and discussed below.


\section{Discussion} 

In this Section, we propose a geometrical interpretation of the results obtained in section 4, 
we then present the mass rate and kinetic power of the outflows of BASS AGN, and finally, we test the outflow fraction as a function of different AGN power tracers.

\subsection{Outflow Fraction}

From the results presented in Fig.~\ref{fig:veloc3}, we observe a clear disparity between type\,1/type\,1.9 and type\,2 AGN:

\begin{itemize}
    \item[-] The occurrence of outflow detections in type\,1/type\,1.9 AGN are twice the one found in type\,2 AGN (i.e. 46\% and 55\% versus 24\%);
    \item[-] Type\,1 and type\,1.9 AGN outflows are almost exclusively blueshifted, while type\,2 AGN exhibit a statistically equal number of redshifted and blueshifted outflows.
\end{itemize}

We interpret this latter difference in the framework of the geometrical unification model of AGN \citep{Antonucci1993,Urry1995}: for type\,1-1.9 AGN we only observe the outflow on the nearside (pointed towards us) because the redshifted outflowing material is obscured for a large range of orientations between the NLR and the host galaxy disk \citep{Fischer2013} and/or the dusty torus, whereas for more edge-on AGN, i.e. type\,2 AGN, we are able to see receding (redshifted) outflows in almost equal probability to blueshifted ones. 

In the following analysis, type\,1.9 and type\,1 are grouped. 
For that, we tested if type\,1.9 can be considered a subcategory of type\,1 or they have most probably a nature of type\,2 AGN. For details, see subsection 5.1.1.

\begin{figure}
\includegraphics[width=\columnwidth]{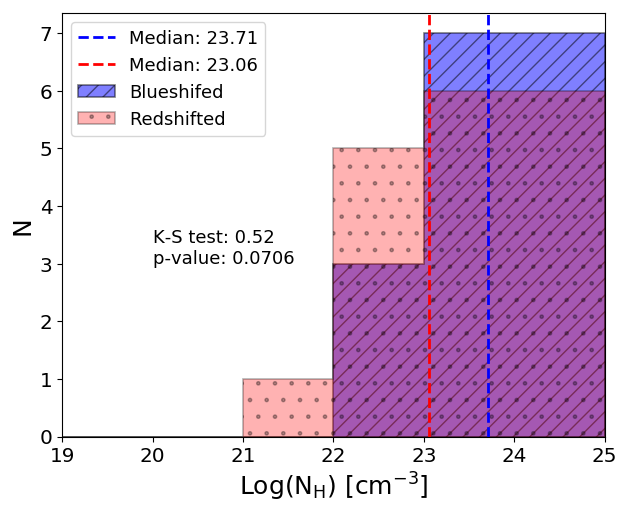}
\caption{Log($N_{\rm H}$) comparing blueshifted and redshifted outflows in type\,2 AGN in fixed bins of column density.}
\label{fig:nh2}
\end{figure}

\begin{figure*}
\centering
\includegraphics[width=0.4\textwidth]{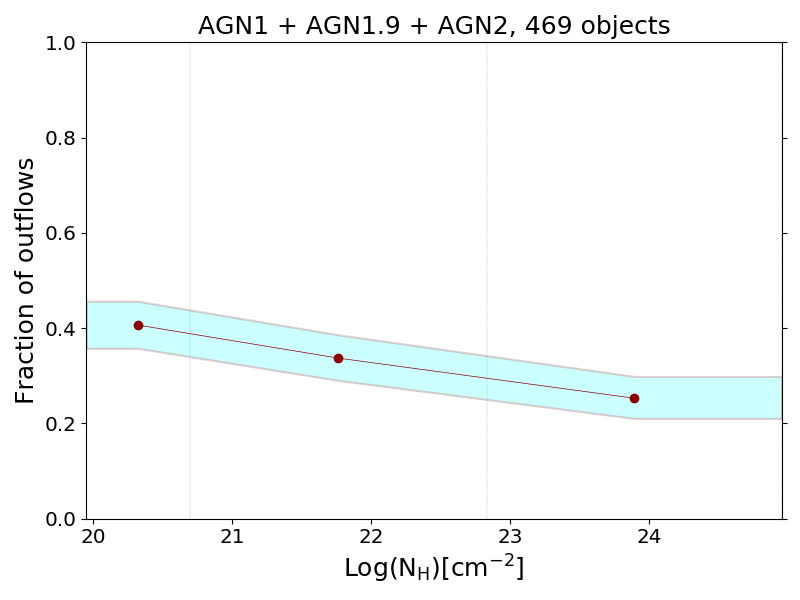}
\includegraphics[width=0.4\textwidth]{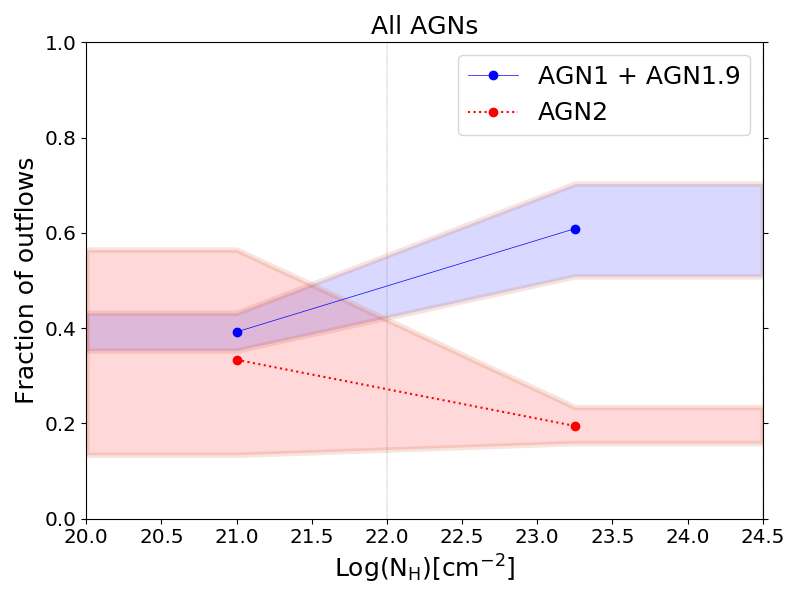}
\caption{Left panel: Fraction of outflows for different bins of $\log{(N_{\rm H})}$ for the full sample. Bins are chosen to have the same number of objects and are indicated with grey vertical lines. Right panel: Fraction of outflows for two different fixed bins of $\log{(N_{\rm H})}$ comparing broad AGN (AGN1+AGN1.9, in cyan) and narrow-line AGN (AGN2, in pink). The shaded area represents the 16th and 84th quantiles of a binomial distribution.}
\label{fig:nh}
\end{figure*}


Our findings are in qualitative agreement with the results presented by \citet{Rakshit2018} for a large sample of low-redshift AGN selected from the SDSS DR12 catalog ($z {<} 0.3$), where blueshifted [O\,{\sc iii}] outflows are more frequently detected than redshifted ones by a factor of 3.6 in type\,1 AGN, while the ratio between blueshifted and redshifted [O\,{\sc iii}] is only 1.08 for type\,2 AGN due to projection and orientation effects. Overall, the outflow fractions found by \citet{Rakshit2018} are larger than the fractions we find here for all AGN types. This is most likely due to the identification and limits imposed by the different outflow detection methods, as well as the inherent [O\,{\sc iii}] luminosity of the different samples. In particular, our thresholds to distinguish winds from gravitational kinematics make our outflow selection criterion more restrictive, in line with the lower fractions found. 


\citet{Perna2017a} selected a sample that includes both type\,1 and type\,2 AGN by cross-matching 2--10\,keV detections from archival XMM-Newton and Chandra data with the SDSS DR12, and adopting the non-parametric method described in Section 4.1. We find similar outflow fractions for the various AGN types as \citet{Perna2017a}. If we further split the fraction of outflows by velocity shift, we find comparable fractions of blue and redshifted outflows in type\,1 AGN, while redshifted outflows are more frequently detected in our work by a factor of 8.

Since the main difference between our sample and that of \citet{Perna2017a} is the selection in the harder (14--195\,keV) versus softer (2--10\,keV) X-ray band, respectively, we wanted to test whether redshifted outflows tend to be at larger $N_{\rm H}$ and therefore could be underrepresented in their sample. Fig.~\ref{fig:nh2} shows that for type\,2 AGN, blueshifted and redshifted outflows populate the same range of $N_{\rm H}$. The probability that the two subsamples come from two distinct families is negligible, with a Kolmogorov-Smirnov test value of 0.52 and p-value ${=} 0.07$. 

In the following subsections, we study how the fraction of outflows is affected (or not) by different AGN properties. 

\subsubsection{\textbf{Type\,2 vs type\,1.9 AGN}}

Regarding the fraction of outflow detections in type\,1/type\,1.9 AGN versus type\,2 AGN, there are a couple of factors that may contribute to the different percentages: a) type\,2 AGN may include a population of extremely obscured objects, whereby an outflow has not yet managed to punch through the obscuring material; b) Type\,2 AGN, on average, will have higher covering factors of the obscuring material around them. Thus, even though they might have intrinsically high $L_{\rm Bol}$ values, most of this luminosity will be reprocessed on small scales, and only a smaller fraction of that energy will be deposited into and drive ionized outflows.

Our results strengthen the hypothesis that type\,1.9 AGN should be considered as a sub-category of type\,1 AGN (e.g. \citealt{HernandezGarcia2017} and references therein). In the following analysis we systematically find that type\,1.9 and type\,1 AGN show similar behaviours, and for this reason we merge these two samples when presenting some results. 

We have also considered the opposite view, i.e. that type\,1.9 behave as type\,2 Seyfert AGN, where the BLR is mimicked in long-slit spectra by an outflow that broadens the emission lines (\citet{Shimizu2018}; hereafter S18). Of the 57 type\,1.9 sources analyzed in S18, 16 (28$\%$) are found to possibly be type\,2 AGN with the BLR simulated by an outflow that is detected in [\rm O\,{\sc iii}]. We crossed-checked our outflow detections with the BASS DR1 sub-sample studied in S18, and found that, in 14 type\,1.9 AGN over 16 sources an outflow is detected based on our criteria. More specifically, we identify 11 blueshifted, 2 redshifted and a symmetric outflow. As an exercise, we moved these 14 type\,1.9 AGN to the type\,2 sub-sample. None of our results are significantly affected by this source redistribution. For example, the outflow fraction of type\,2 AGN increases from 24$\%$ to 29$\%$, split into 17$\%$ approaching and 11$\%$ receding outflows.

\subsection{Outflows Fraction and Gas Column Density} 

\begin{figure}
\includegraphics[width=\columnwidth]{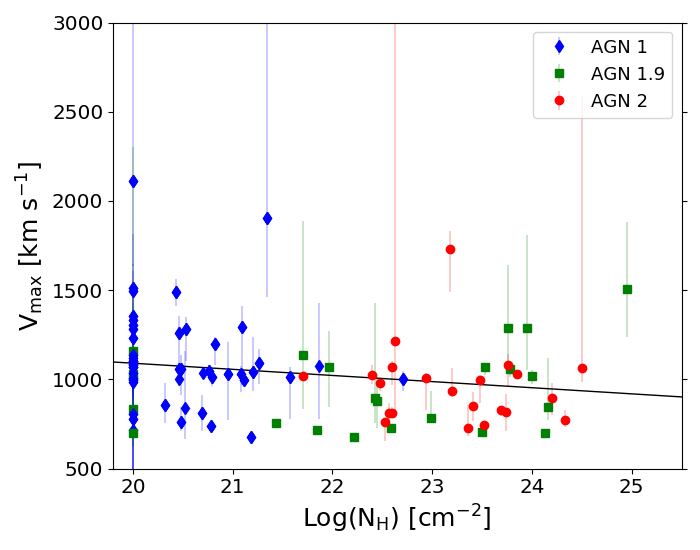}
\caption{Outflow velocities versus  $\log{(N_{\rm H})}$ for the different AGN types.}
\label{fig:nh3}
\end{figure}

Theoretical models in which AGN activity is triggered by galaxy mergers propose an evolutionary path whereby obscured AGN reside in star-forming galaxies during a period of rapid SMBH and galaxy growth, followed by a period where the AGN drives outflows that expel the surrounding material and reveal an unobscured AGN (e.g., \cite{Hopkins2006, Hopkins2008}). As argued by past studies (e.g., \cite{Harrison2016}), under this scenario we might expect a larger outflow fraction among the most X-ray obscured AGN [$\log{(N_{\rm H})} {>} 22$\,cm$^{-2}$]. Using the column density ($N_{\rm H}$) estimated from X-ray spectral analysis of \citet{Ricci2017d} we observe the opposite trend (see Fig. ~\ref{fig:nh}, left panel). In addition, Fig.~\ref{fig:nh3} shows that higher velocities are not associated with the most obscured sources, where the data are dispersed and do not correlate significantly according to the Pearson test and Spearman rank (p-value ${=} 0.05$, $R^{2} {=} 0.04$ and p-value ${=} 0.01$, $\rho {=} {-0.27}$, respectively). This is consistent with past findings (e.g., \cite{Harrison2016}).
At $\log{(N_{\rm H})} {<}  21$\,cm$^{-2}$, indeed the fraction of detected outflows is almost 2 times higher than in more absorbed sources (a 2-$\sigma$ difference). 
However, projection effects might result in underestimated velocities for type\,2 AGN relative to type\,1. Furthermore, contamination from the host galaxy might prevent the measurement of the highest velocities, which correspond to the faintest wings of the [O\,{\sc iii}] line profiles. Our results are in agreement with the analysis of \citet{Harrison2016}, who analyzed a sample of hard X-ray selected AGN at $0.6{<}z{<}1.7$. To this end, we first highlight the left panel of Fig.~\ref{fig:nh}, where a clear anti-correlation between of the outflow fraction and $N_{\rm H}$ can be seen; 
The right panel of Fig.~\ref{fig:nh} disentangles the evolution of type\,1+1.9 and type\,2 AGN, demonstrating that the outflow fractions for type\,1 and type\,1.9 actually increase with $N_{\rm H}$. For type\,2 AGN, we observe the opposite trend, but with lower significance. We note that this type\,1+1.9 AGN trend in Fig.~\ref{fig:nh} effectively corresponds to the transition from a population of type\,1 AGN with low $N_{\rm H}$ values to a population of type\,1.9 AGN with large $N_{\rm H}$ values.

\subsection{Outflow Fraction and Luminosities}

\begin{figure}
\includegraphics[width=\columnwidth]{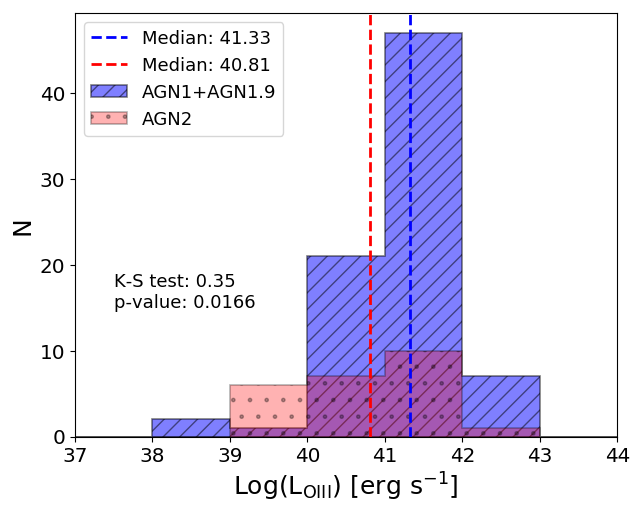}
\caption{Luminosity distribution of the sources with an outflow detection. type\,1+type\,1.9 are in blue and type\,2, in red. We consider here the narrow component of the [O\,{\sc iii}] line.}
\label{fig:LOIII1}
\end{figure}

Fig.~\ref{fig:LOIII1} shows the distribution of the AGN [O\,{\sc iii}] luminosities with an outflow detection for fixed bins of luminosity, where the [O\,{\sc iii}] luminosities correspond to the fitted narrow emission components, representing the AGN. We can see that on average type\,1+type\,1.9 are present at higher [O\,{\sc iii}] luminosities than type\,2. This result is to be expected, since type\,2 have typically lower luminosities, and [O\,{\sc iii}] scales with the X-ray luminosity. 

\begin{figure}
\includegraphics[width=\columnwidth]{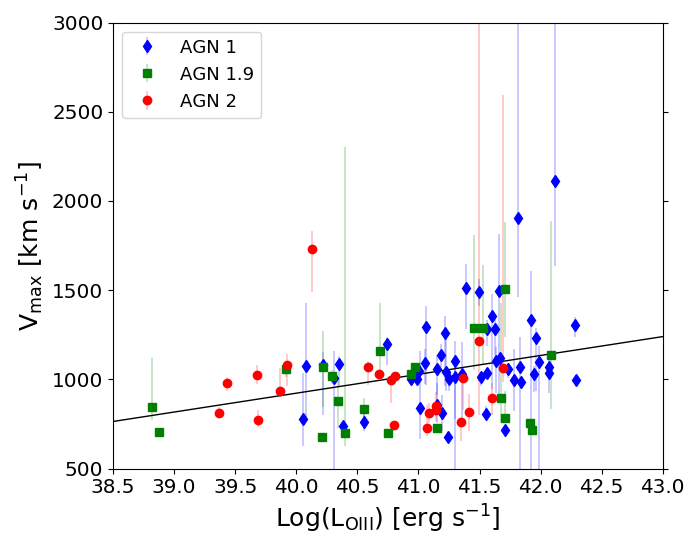}
\caption{Outflow velocities as a function of $\log{L_{\rm [O\,III]}}$ for the different AGN types. We consider here the narrow component of the [O\,{\sc iii}] line.}
\label{fig:LOIII2}
\end{figure}

Fig.~\ref{fig:LOIII2} shows that the highest maximum velocities (${>} 1200$\,km\,s$^{-1}$) are detected almost exclusively for [O\,{\sc iii}] luminosities higher than $10^{41}$\,erg\,s$^{-1}$. This is in agreement with previous works where a slightly positive trend between the outflow velocity and the AGN luminosity was found \citep{Reyes2008, Yuan2016, Perna2017a, Fiore2017,Rakshit2018}. However, there is no strong correlation between outflow velocity and [O\,{\sc iii}] luminosity. We use Spearman rank and Pearson test to quantify a possible correlation. The correlation coefficients are 0.3 and 0.1, with probabilities of ${<} 0.001$ for the correlation being observed by chance, respectively. \citet{Perna2017a} finds a positive trend, although we note that this result included AGN with low velocity kinematics ($\rm v_{max} {<} 650$\,km\,s$^{-1}$), which are excluded in our analysis. We also tested how the outflow velocity relates with bolometric luminosity and we found no correlation between both quantities, while e.g. \citet{Fiore2017} found a positive trend. However, we note that we cover a smaller range of luminosities than \citet{Fiore2017}, between ${\sim} 10^{44}$--$10^{46}$\,erg\,s$^{-1}$ versus  ${\sim}10^{43}$--$10^{48}$\,erg\,s$^{-1}$.

\begin{figure}
\includegraphics[width=\columnwidth]{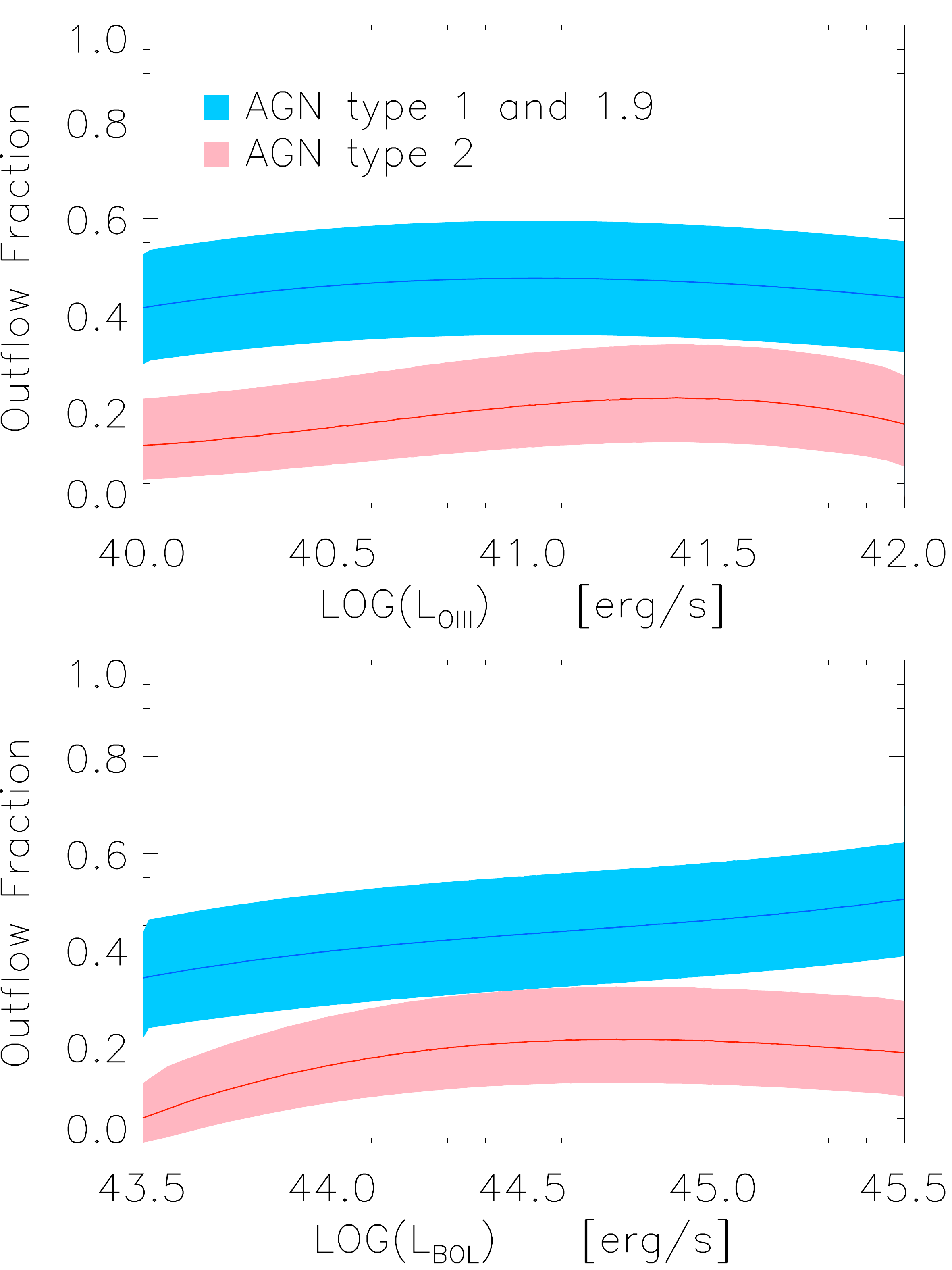}
\caption{Outflow fraction as a function of luminosities $\log{L_{\rm [O\,III]}}$ and AGN $\log{L_{\rm Bol}}$ for the different AGN types. Fractions were obtained considering the sample of a fixed number of neighbouring objects in term of luminosity. Then, the resulting curve was then fitted with a low order polynomial.}
\label{fig:frac_lum}
\end{figure}

On the other hand, when we compare in Fig.~\ref{fig:frac_lum} the fraction of outflows to $L_{\rm [O\,III]}$ and $L_{\rm Bol}$, we note that type\,1+type\,1.9 show about $20\%$ larger outflow fraction than type\,2 for the different ranges of luminosities, and with little to no dependence on AGN luminosity for either type\,1+type\,1.9 or type\,2 AGN.

\begin{figure}
\includegraphics[width=\columnwidth]{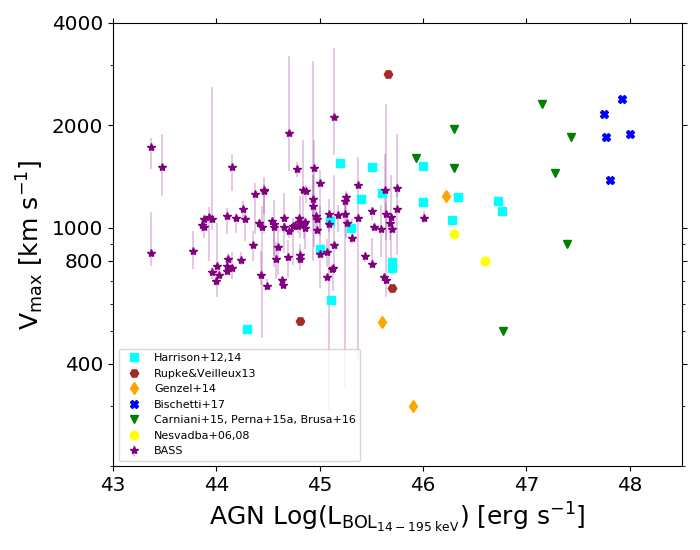}
\caption{Maximum outflow velocities as a function of AGN $\log{(L_{\rm Bol})}$ compared with several literature samples \citep{Bischetti2017, Carniani2015, Perna2015a, Brusa2016, Genzel2014, Harrison2012,Harrison2014, Rupke2013, Nesvadba2006,Nesvadba2008}.}
\label{fig:V_Lbol_lit}
\end{figure}

Fig.~\ref{fig:V_Lbol_lit} shows the estimated $\rm v_{max}$ as a function of $L_{\rm Bol}$. We compare our findings with ionized outflows of: obscured X-ray selected quasars \citep{Brusa2016, Perna2015a}, [OIII]-loud quasars at $z {\sim} 1.5$--2.5 with $L_{\rm Bol} {>} 10^{47}$\,erg\,s$^{-1}$ \citep{Carniani2015}; massive AGN at $z {\sim} 2$ (Genzel et al. 2014); low and high z (mostly type\,2) AGN \citep{Harrison2012,Harrison2014}; and high z radio-galaxies \citep{Nesvadba2006, Nesvadba2008}. While when considering the BASS sample alone, no correlation is detected, the extension of the sample to larger AGN luminosities seems to imply  mild positive correlation between maximum outflow velocity and AGN luminosity, with the ionized outflows discovered in BASS covering low-to-moderate velocities at the low AGN luminosity end of the diagram and various literature samples covering moderate-to-high AGN luminosities. We caution anyhow that the various sample selections applied in literature might affect the result. 

\subsection{Trend with the Accretion Rate}

\begin{figure}
\includegraphics[width=\columnwidth]{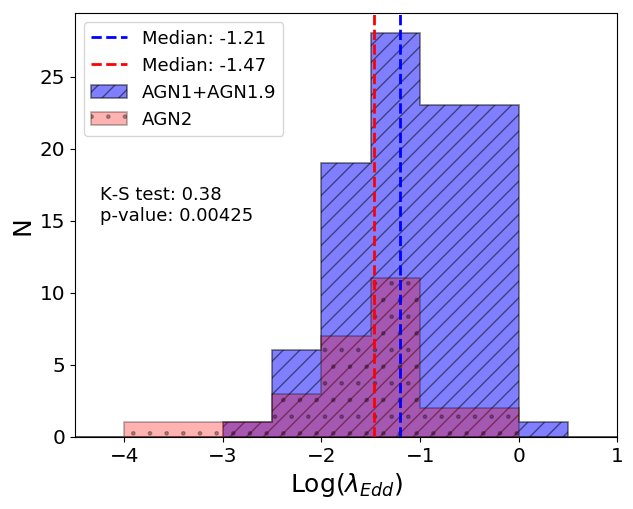}
\caption{Distributions of Eddington ratio $\lambda_{\rm Edd}$ for broad-line (type\,1+type\,1.9, dashed blue line) and narrow-line AGN (type\,2, dotted red line) with outflow detections. Broad-line AGN have higher average $\lambda_{\rm Edd}$ values than narrow-line AGN.}
\label{fig:edd1}
\end{figure}

When we compare the distribution of outflow detections with the accretion rate ($\lambda _{Edd}$), we find that type\,1 AGN with outflows have a higher Eddington ratio than type\,2 AGN with outflows (See Fig.~\ref{fig:edd1}, with fixed bins of $\lambda _{Edd}$). However, this is simply reflecting the Eddington ratio difference found for the general population of type\,1 and type\,2 AGN (e.g., \cite{Trump2011}). 

\begin{figure*}
\centering
\includegraphics[width=0.3\textwidth]{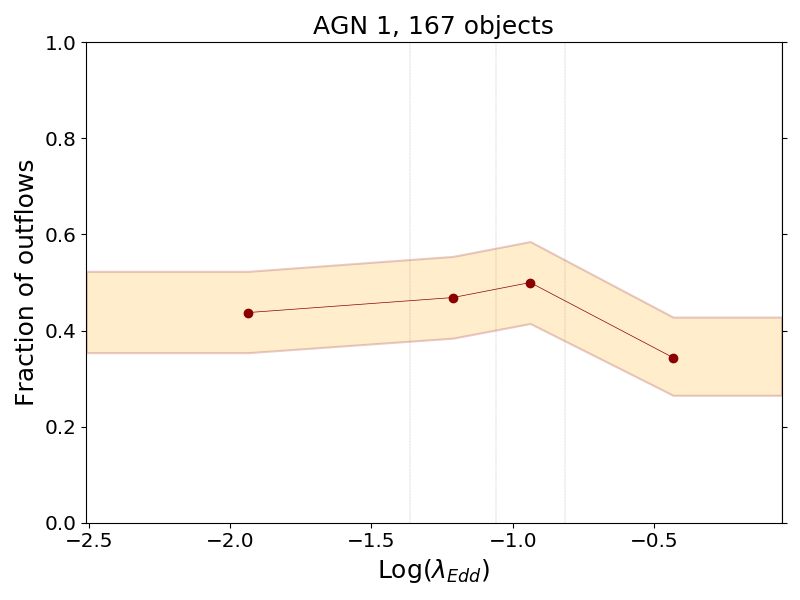}
\includegraphics[width=0.3\textwidth]{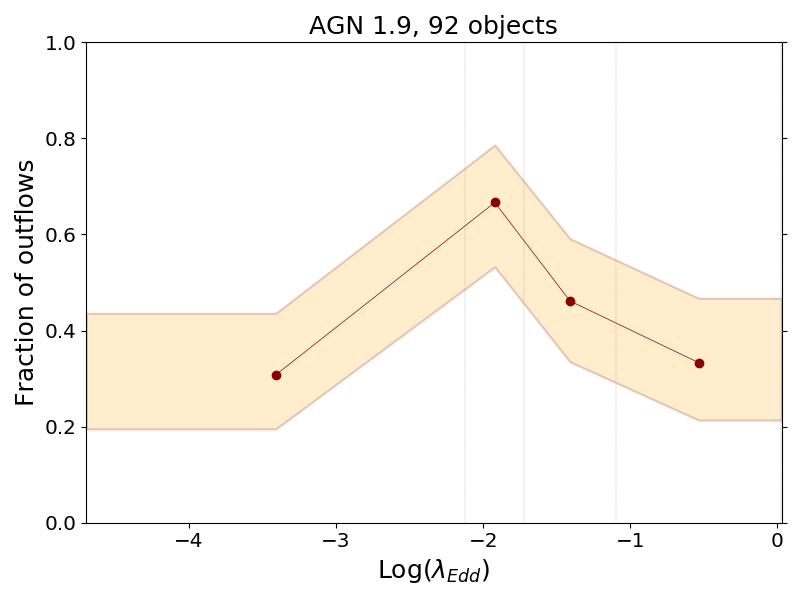}
\includegraphics[width=0.3\textwidth]{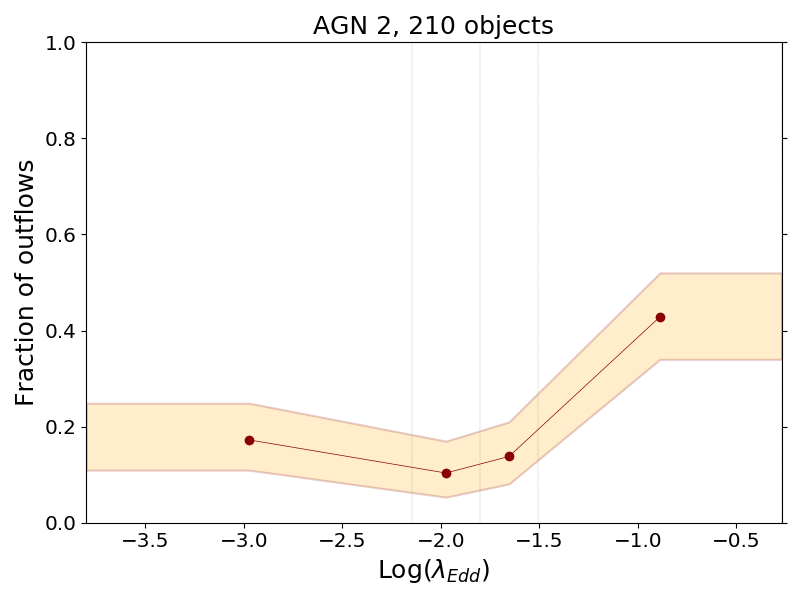}
\caption{Fraction of detected outflows as a function of Eddington ratio $\lambda_{\rm Edd}$, with each panel tracing the accretion rate trend for a given AGN type. Shaded areas represent the 16th and 84th quantiles of a binomial distribution, where bins are chosen to have the same number of objects and are indicated with vertical grey lines.
}
\label{fig:edd2}
\end{figure*}

From Fig.~\ref{fig:edd2}, we can see that the outflow fraction in type\,1 and type\,1.9 AGN is higher than in type\,2 ($\sim$ 50\% versus 24\%). We can explain this difference with a couple of factors that can contribute: a) type\,2 AGN include a population of highly obscured objects, where outflow have not yet managed to punch through the obscuring shell and drive the outflows; b) type\,2 AGN, on average, will have higher covering factors of the obscuring material around them. Thus, even though they might have intrinsically high $\rm L_{BOL}$, most of it will be reprocessed on small scales (i.e. most of the turbulent gas is contained within the torus), and only a smaller fraction of that energy will be deposited into ionizing and driving the outflows.

More intriguing is the fact that the outflow fraction appears to depend on the Eddington ratio: wee see that the outflow fraction for type\,2 AGN increases with $\lambda _{\rm Edd}$, while the trend for type\,1 and type\,1.9 AGN seems to decrease or remain flat with accretion rates (See Fig.~\ref{fig:edd2}). 


\begin{figure}
\includegraphics[width=\columnwidth]{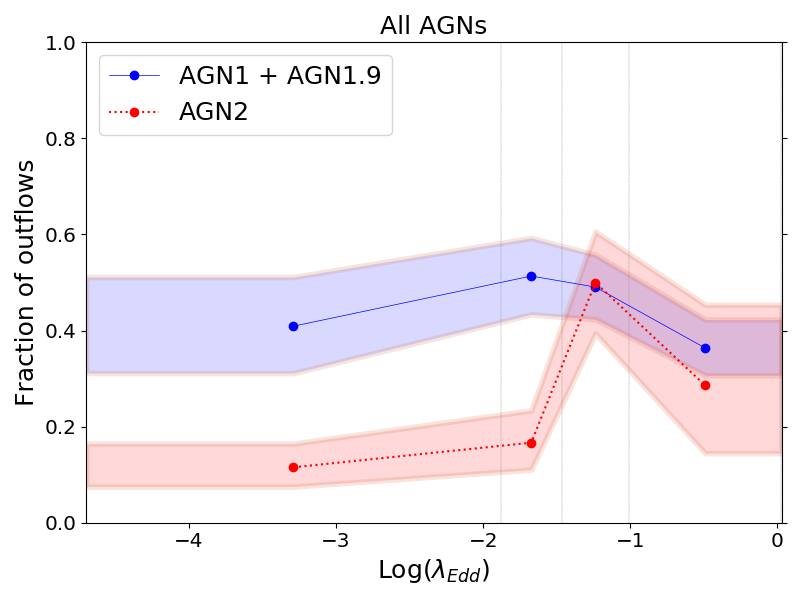}
\caption{Fraction of outflows as a function of Eddington ratio $\lambda_{\rm Edd}$ comparing broad-line AGN (AGN1+AGN1.9; blue) and narrow-line AGN (AGN2; red). Shaded areas represent the 16th and 84th quantiles of a binomial distribution, where bins are chosen to have the same number of objects and are indicated with grey vertical lines. Broad-line AGN have a high and relatively flat fraction of outflow detected at different $\lambda_{\rm Edd}$, while AGN2 have a very low detection fraction at low $\lambda_{\rm Edd}$ but increase dramatically to similar values as broad-line AGN around $\log{(\lambda_{\rm Edd})} {\gtrsim} {-}1.5$.}
\label{fig:edd3}
\end{figure}


This behavior is even clearer when we merge type\,1 and type\,1.9 AGN together and compare their trend as a function of $\lambda_{\rm Edd}$ with respect to type\,2 AGN (See Fig.~\ref{fig:edd3}). While for the former there is no trend, the fraction of outflows in type\,2 AGN increases significantly (${\sim} 3\sigma$) above $\log{(\lambda_{\rm Edd})} {\gtrsim} {-}1.5$. 


\begin{figure}
\includegraphics[width=\columnwidth]{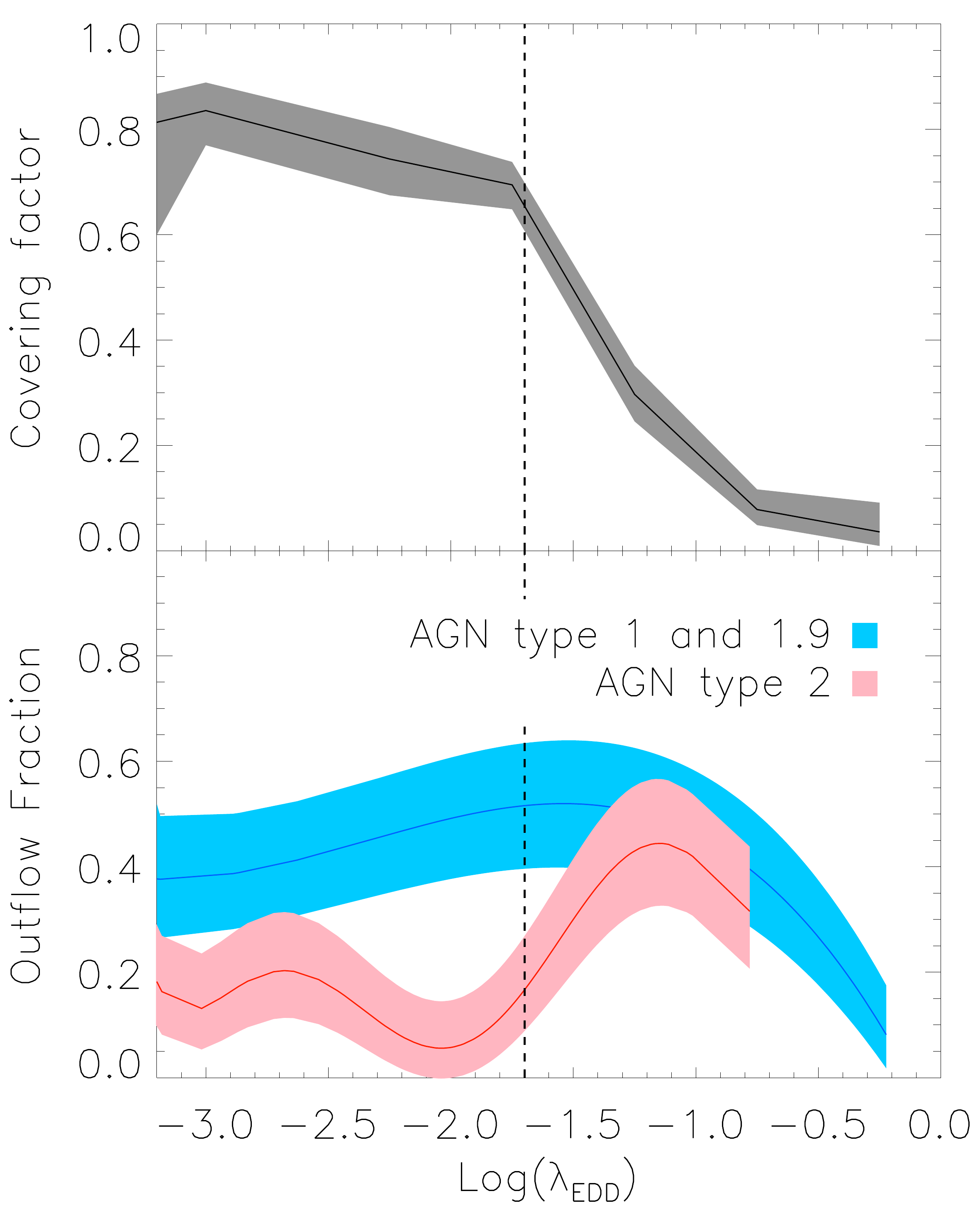}
\caption{Top panel: Fraction of obscured sources as in \citet{Ricci2017c} as a function of Eddington ratio for the BASS Sample. Bottom panel: Outflow fraction of type\,1+type\,1.9 (blue) and type\,2 AGN (red) as a function of $\lambda_{\rm Edd}$. These fractions were obtained considering the sample of a fixed number of neighbouring objects in term of Eddington ratio. The resulting curves (blue and red) were then fitted with a low order polynomial.}
\label{fig:edd_Claudio}
\end{figure}

An interesting result to stress is shown in Fig.~\ref{fig:edd_Claudio}: In the uper panel, the behavior between type\,1+type\,1.9 and type\,2 AGN start to be significantly different at a given Eddington ratio, that is for $\log{(\lambda_{\rm Edd})} {>} {-}1.7$. This point is the same Eddington ratio where the covering factor of obscured sources decreases significantly with the Eddington ratio according to \citet{Ricci2017c}. In that scenario, this value can be interpreted as a threshold above which radiation pressure on dusty gas is able to create outflows \citep{Fabian2016, Ishibashi2018}. Then, when the Eddington ratio reaches the highest values (${\gtrsim} {-}1.2$), most of the material around the BH has been blown away and winds cannot be sustained efficiently anymore. 
We compared our data with such interpretation, and can see that indeed the fraction of outflows increases for type\,2 at $\log{(\lambda_{\rm Edd})} {\gtrsim} {-}1.7$ (see Fig.~\ref{fig:edd_Claudio}, bottom panel, red curve) to then drop at $\log{(\lambda_{\rm Edd})} {\gtrsim} {-}0.8$. Such drop can be understood as a selection effect: small covering factors mean it is very unlikely that these sources will be observed as type\,2 AGN. However, we do not see a similar trend for type\,1+type\,1.9. In fact, puzzlingly, it seems that the outflow fraction in both type\,1+1.9 and type\,2 AGN starts to decrease above Eddington ratios of $\log{(\lambda_{\rm Edd})} {\gtrsim} {-}1.2$. At high $\log{(\lambda_{\rm Edd})}$, the covering fraction of the obscuring material is rather low, as it can be seen in the upper panel. Thus, in type\,2 AGN the rise of outflow fraction happening at ${\lesssim} {-}1.7\log{(\lambda_{\rm Edd})}{\lesssim} {-}0.8$ can be interpreted as the condition at which the obscuring material surrounding the engine, starts to blow out. Such effect is not visible in type\,1 AGN because they are not characterized by high covering factors. This is confirmed by the fact that above $\log{(\lambda_{\rm Edd})}\sim -1.0$ the blue (type\,1 AGN) and red (type\,2 AGN) curves in the bottom panel of Fig.~\ref{fig:edd_Claudio} are again consistent, i.e. once the Eddington ratio becomes such that outflows cannot be sustained anymore. Our interpretation is supported by a Fisher's exact test to calculate the \textit{p}-value significance of the difference in the two samples proportions as defined by the Eddington ratio of $\log{(\lambda_{\rm Edd})} = {-}1.7$. When comparing the outflows fractions for type\,1+type\,1.9 and type\,2 AGN we obtained \textit{p}-value = 0.001 for low $\log{(\lambda_{\rm Edd})}$ and \textit{p}-value = 0.12 for high $\log{(\lambda_{\rm Edd})}$.

\subsection{Outflow Kinematics}

In order to quantify the impact of these ionized outflows on the host galaxy, we need to determine their spatial extension and energetic. In this section, we calculate the mass outflow rate [$\dot{M}_{\rm out}$ ($M_{\rm \odot}$\,yr$^{-1}$)] and the kinetic power [$\dot{E}_{\rm kin}$ (erg\,s$^{-1}$)] of the ionized outflows in our sample. 

Considering the simple model of a spherically/biconically symmetric mass-conserving free wind, we can estimate the outflowing ionized gas mass, $\dot{M}_{\rm out}$, from the fluid field continuity equation, assuming that most of the oxygen consists of $O^{2+}$ ions with a gas temperature of $T {=} 10^{4}$\,K (typical temperature measured for the NLR), as in \citet{Carniani2015}.

If the mean density of an outflow covering the solid angle $\Omega \pi$ is given by $\rho {=} \frac{3M^{out} _{ion}}{\Omega \pi R^3}$, then $\dot{M}_{\rm out}$ can be estimated locally at a given radius $r$ (e.g., \cite{Feruglio2015}) by:

\begin{equation}
\dot{M}_{\rm out} = \Omega \pi r^{2} \rho {\rm v} = 3\frac{M^{\rm out} _{\rm ion} {\rm v}}{r}
\label{eq:Mout}
\end{equation} \\

\noindent where v is v$_{\rm max}$, given by Eq.\,(2) (assumed constant with radius and spherically symmetric in this simple bi-conical model) and $M^{\rm out}_{\rm ion}$ is the ionized outflowing gas mass given by:

\begin{equation}
M^{\rm out} _{\rm ion} = 4.0\times 10^7 M_{\odot}\left( \frac{C}{10^{[O/H]}}\right) \left( \frac{L_{\rm [OIII]}}{10^{44}\,{\rm erg\,s^{-1}}} \right) \left( \frac{<n_{\rm e}>}{10^3\,{\rm cm}^{-3}} \right)^{-1}
\label{eq:Mion}
\end{equation} \\

\noindent where $L_{\rm [OIII]}$ is the luminosity of the [O\,{\sc iii}]$\lambda 5007$ line tracing the outflow (from the flux of the broad line component), \hbox{$C {=} <n_{\rm e}>^{2}/<n_{\rm e} ^2>$} and $n_{\rm e}$ is the electron density. We assume $C {\sim} 1$ based on the hypothesis that all the ionized gas clouds have the same density, and log([O/H])${\sim} 0$ (solar metallicity). However, large uncertainties in $n_{\rm e}$ and $r$ propagate to a large uncertainty in the ionized mass outflow rate estimate, up to an order of magnitude. Indeed, the measured gas density $n_{\rm e}$ varies from a few hundreds up to several thousands of cm$^{-3}$ for different methods in different samples \citep{Kakkad2018, Fiore2017, Bischetti2017, Kakkad2016, Carniani2015, Perna2015a, Perna2015b, Harrison2014, Canodiaz2012, BaronNetzer2019}. In addition, we do not know the physical extent of our outflows because we are limited by single-slit spectroscopic observations with no spatial information. 

Therefore, in order to limit the effects of such uncertainties, we use the correlation between the outflow size $r$ ($R_{\rm out}$) and the luminosity ($L_{\rm [O\,III]}$) recently found by \citet{Kang2018} to estimate $R_{\rm out}$: 

\begin{equation}
\log{(R_{\rm out})} = 0.28\times \log{(L_{\rm [O\,III]})} - 11.34.
\label{eq:Rout}
\end{equation} 
 
To estimate the electron density we refer to \citet{BaronNetzer2019}, where the authors use optical line ratios of [OIII]/$\rm H_{\beta}$ and [NII]/$\rm H_{\alpha}$, and the location of the wind estimated from mid-infrarred emission. They found a value of $\rm n_{e}\sim 10^{4.5}\ cm^{-3}$, suggesting that the [SII]-based method commonly used by several authors, underestimates the true electron density in the outflowing gas by roughly 2 orders of magnitude.

Then, we derive the kinetic power associated to the outflow from the mass outflow rate by:

\begin{equation}
\dot{E}_{kin} = \frac{1}{2} \dot{M}_{out} v^2
\label{eq:Ekin}
\end{equation} 

We estimate the wind momentum load as:

\begin{equation}
w = \frac{\dot{P}_{\rm out}}{\dot{P}_{\rm AGN}} = \frac{\dot{M}_{\rm out}{\rm v}}{L_{\rm Bol}/c} 
\label{eq:Pout}
\end{equation} \\

\noindent where $\dot{M}_{\rm out}$v is the outflow momentum rate, and $L_{\rm Bol}$/$c$ is the AGN radiation pressure momentum. If $w {<} 10$, outflows are considered momentum-conserving, i.e. while expanding the gas decreases in temperature and releases energy through radiation. If $w$ is larger, the outflows are considered energy-conserving, i.e. they expand adiabatically with constant temperature (see \cite{Zubovas2014} for details). 

\begin{figure}
\includegraphics[width=\columnwidth]{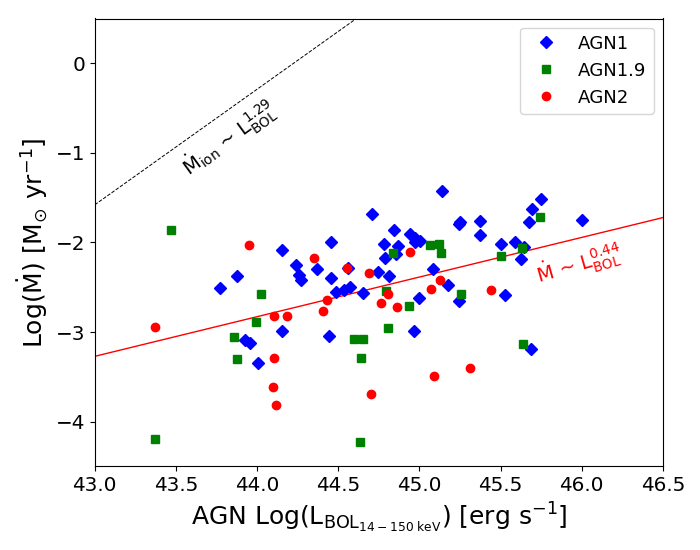}
\caption{Mass outflow rate of the ionized outflows as a function of $L_{\rm Bol}$. Dotted black line corresponds to the best-fit relation derived for ionized outflows of \citet{Fiore2017}, and solid red line corresponds to our best fit.}
\label{fig:Mdot}
\end{figure}

\begin{figure}
\includegraphics[width=\columnwidth]{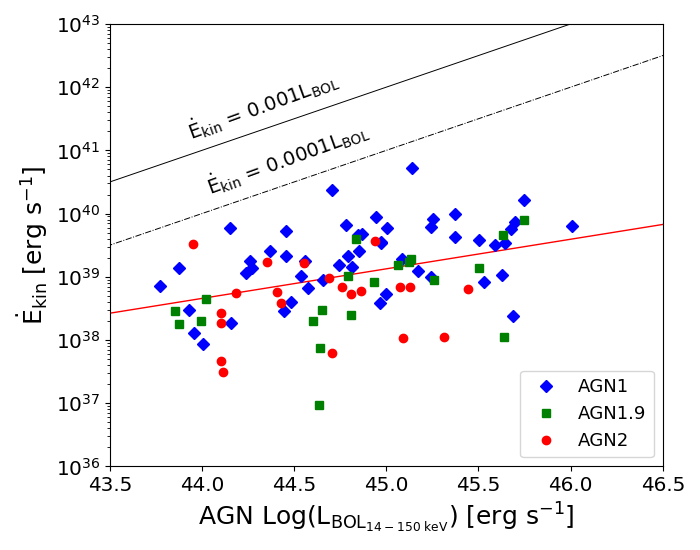}
\caption{Kinetic power of the ionized outflows as a function of $L_{\rm Bol}$. Dotted-dashed and solid lines represent an outflow kinetic power that is 0.1\% and 0.01\% of the AGN bolometric luminosity, respectively.}
\label{fig:Ekin}
\end{figure}

The estimated values of the mass outflow rate and kinetic energy as a function of $L_{\rm Bol}$ (as estimated from the hard X-ray luminosity) for our sample are presented in Figs.~\ref{fig:Mdot} and \ref{fig:Ekin}. We find that the energetics of the gas outflows are mildly correlated with bolometric luminosity. We use the Spearman rank and Pearson correlation test to derive the significance of the observed trends: we find coefficients of 0.46 and 0.20, with probabilities of ${<} 0.001$ that the correlation is observed by chance. According to Fig.~\ref{fig:Mdot}, ionized outflows at higher luminosities appear to expel a larger amount of the total ionized outflowing gas than outflows at lower luminosities. In addition, we note that they lie below the correlation found for ionized winds by Fiore et al. 2017, where the correlation between $\dot{M}$ and $L_{\rm Bol}$ has a log linear slope of 1.29 $\pm$ 0.38; instead, we find a slope of 0.44. This could be due to a different ionization density and spatial extension of the outflow region. The correlation of our dataset is not as steep as \citet{Fiore2017}. One explanation may be the different luminosity range covered by the BASS sample. Our sources are fainter than the sample presented in \citet{Fiore2017}, and the onset of the correlation between outflow energetics and AGN bolometric luminosity could happen at higher luminosities.

Fig.~\ref{fig:Ekin} shows that the average kinetic power of our sample is less than 0.0001\% $\rm L_{Bol}$, indicating very low energy conversion efficiencies, which are lower than some results in the literature. One explanation here may be the different values assumed for the gas properties to estimate the kinetic power. For example, \citet{Rakshit2018} found that their sample has on average $\sim$0.001\% $L_{\rm Bol}$, assuming an electron density of $n_{\rm e}$ = 272 cm$^{\rm -3}$, while we used \textbf{$n_{\rm e}$ = $10^{4.5}$ cm$^{\rm -3}$}. However, we found that our values of mass outflow rate and kinetic energy are in agreement with authors using higher values of electron density (e.g. \cite{BaronNetzer2019}). This reinforce the fact that using different assumptions in geometry and intrinsic properties of the wind leads important differences when we estimate their mass and kinematic power.

On the other hand, we estimate low wind-momentum loads for all AGN types, $<$ 0.1, in agreement with \citet{Fiore2017} where the range is estimated to be between 0.01 and 30. This suggests that the BASS AGN winds are probably momentum-conserving, as predicted by the \citet{King2003} model.

Finally, we would like to stress that we used the $L_{\rm [O\,III]}$ of the outflow component to estimate $\dot{M}_{\rm out}$ and $\dot{E}_{kin}$, so we do not expect to the perceived correlations showed in Figs.~\ref{fig:Mdot} and \ref{fig:Ekin} are due to a trend between $L_{\rm [O\,III]}$ and $L_{\rm Bol}$. Moreover, the total $L_{\rm [O\,III]}$ and bolometric luminosity quantities are found to be not correlated, with a huge scatter between $L_{\rm [O\,III]}$ and X-ray ($\sim L_{\rm Bol}$) emission (see e.g. \citet{Berney2015, Ueda2015}).

\section{Summary and Conclusions}

The aim of this work is to test the incidence of ionized outflows for a large sample of nearby hard X-ray selected AGN, and to study how the outflow properties are related to different AGN tracers (X-ray, [O\,{\sc iii}], and bolometric luminosities, $M_{\rm BH}$, $\lambda_{\rm Edd}$). 


The modelling of the optical spectra of hard X-ray selected AGN allowed us to derive the incidence and properties of ionized outflows in an unbiased/complete sample covering a wide range of AGN bolometric luminosities, and to study the differences between obscured and unobscured AGN. 

According to this, to investigate the presence of an outflow, we focus on the [O\,{\sc iii}]$\lambda$4959,5007 emission lines. In particular, we used a multi-component fitting procedure to account for the faint wings of [O\,{\sc iii}] associated with an outflow signature. Outflow velocities were estimated using two different approaches: following the criteria of \citet{Rupke2013} who use the parameters of both fitted components of the line, and a non-parametric method that is expected to be less sensitive at low S/N. 

We found that 38$\pm$2\% of our AGN sample analysed (178/469) present detected outflows, mostly blueshifted, and that the fraction of blue vs redshifted outflows in our sample is consistent with a simple geometrical unification of Type1/type\,2 AGN. 

We test how the outflow fraction and velocities relate to the AGN properties. We observe an increasing outflows fraction as a function of Eddington ratio for type\,2 AGN, and we find weak trends between outflow velocity and AGN luminosity (as traced by $L_{\rm X}$, $L_{\rm O\,{\sc iii}}$, or $L_{\rm Bol}$), and no evident trend between outflow incidence and X-ray obscuration. The Eddington ratio seems to be a fundamental parameter to understand the type\,1 vs type\,2 AGN dichotomy from the point of view of outflow frequency.

Finally, we estimate the kinetic energy and power of the outflows, adopting several assumptions about the physical geometry and gas conditions. An important caveat to bear in mind is that we are only able to trace the ionized phase of the outflows, and we must factor in the neutral and molecular components (which will require additional observations), in order to put constraints on the overall mechanics governing outflows in these sources.

\section*{Acknowledgements}

We are grateful to Prof. M. Elitzur for fruitful discussion on outflow statistics.
We acknowledge support from the Ministry of Education, Science, and Technological Development of the Republic of Serbia through the projects Astrophysical Spectroscopy of Extragalactic Objects \#176001 (MS) and Gravitation and the Large Scale Structure of the Universe \#176003 (MS);
CONICYT grants Basal-CATA Basal AFB-170002 (FEB, FR), FONDECYT Postdoctorado 3180506 (FR); the Ministry of Economy, Development and Tourism's Millennium Science Initiative through grant IC120009, awarded to The Millennium Institute of Astrophysics, MAS (FEB); NASA through ADAP award NNH16CT03C (MK). KO acknowledges support from the Japan Society for the Promotion of Science (JSPS, ID: 17321).





\bibliographystyle{mnras}
\bibliography{bibitextFile}

\bsp	
\label{lastpage}
\end{document}